\documentclass[12pt]{article}
\usepackage{epsfig}
 
\def\pmb#1{\setbox0=\hbox{#1}    
 \kern-.025em\copy0\kern-\wd0    
 \kern.05em\copy0\kern-\wd0      %
 \kern-.025em\raise.0433em\box0} %
\def\sgma{\pmb{$\sigma$}}        %
\def\nbla{\pmb{$\nabla$}}        %
\font\amsfnt=msam10 
\begin{document}
\date{}
\title{
Generalization of the effective mass method for semiconductor structures
with atomically sharp heterojunctions}
\author{
E. E. Takhtamirov$^{\dagger}$ and V. A. Volkov$^{\ddagger}$
\\
Institute of Radio Engineering and Electronics of RAS
\\
Mokhovaya 11, 101999 Moscow, Russia
\\
$^{\dagger }$e-mail: takhtam@cplire.ru
\\
$^{\ddagger }$e-mail: VoVA@cplire.ru
}
\maketitle

\begin{abstract}
The Kohn-Luttinger envelope-function method is generalized to the case of heterostructures
with atomically sharp heterojunctions based on lattice-matched layers of related
semiconductors with zinc-blende symmetry. For electron states near the $\rm \Gamma $
point in (001) heterostructures the single-band effective-mass equation is derived,
taking into account both the spatial dependence of the effective mass and effects
associated with the atomically sharp heterojunctions. A small parameter is identified,
in powers of which it is possible to classify the various contributions to this equation.
For hole states only the main contributions to the effective Hamiltonian, due to the
sharpness of the heterojunctions, are taken into account. An expression is derived for
the parameter governing mixing of states of heavy and light holes at the center of the
2D Brillouin zone.
\end{abstract}
\vspace{0.3cm}

\section{Introduction}

The Kohn-Luttinger effective-mass method \cite{kohn-lut,lut} is widely
used to describe electron states in semiconductors when applied are
external fields varying smoothly over scales of the lattice constant $a$.
Although the original method, based on the formalism of envelope functions,
is applicable only to homogeneous semiconductors, various modifications of it
have been used to describe the electron states in semiconductor heterostructures.
In recent years there has been a revival of discussion on the applicability
of the effective-mass method to describe electron and hole states in real
nanostructures \cite{leibler}-\cite{kisin}. Many different modifications
of the effective-mass method have been proposed, which apply to the case of a
spatially varying effective mass $m ( {\bf r})$. There are two ways of
constructing the effective-mass approximation for heterostructures.
1) Derivation of the effective Hamiltonian for the envelope functions, defined over
all space. By integrating the effective-mass equation, which
contains this Hamiltonian, near the heteroboundary it is possible
to obtain boundary conditions on the envelope functions if needed.
2) Derivation or, as is done much more often, postulation of phenomenological
boundary conditions on the envelope functions at the heteroboundary.
This approach makes use of symmetry arguments, continuity of the
probability flux density, etc. But these arguments, as a rule, are
insufficient to uniquely determine the boundary conditions. The second approach is
applicable in the case of sharp heterojunctions; all the models in which such boundary
conditions were obtained dealt with mathematically sharp heterojunctions between the
left-hand and right-hand materials. It is implicitly assumed that the envelope function
on the left side (on the right side) of the heteroboundary
satisfies the same equation as in the bulk case for the left-hand
(right-hand) material. In this case the very delicate problem of the
accuracy excess arises, which, by the way, has not been discussed to this date:
the boundary conditions should hold with the same accuracy as the equations for the
envelope functions hold.

Below we will follow the first approach, in which it is possible to rigorously treat
the problem of accuracy, see Sec.~3.

It is well known that two main problems arise along the path of constructing a
common equation for the envelope functions. The first is the problem of ordering
the momentum operators in the kinetic-energy operator (due to the non-commutativity
of the momentum operator and the function $m ( {\bf r})$), on the form of which
the solutions of the effective-mass equations can depend substantially \cite{brezini}.
The second problem is that the effective potential near a heteroboundary, as a rule,
is not a smooth function on scales of the order of $a$. This
calls into question even the validity of using differential equations
in the method of envelope functions. Let us discuss these problems in the indicated order.

\subsection{Account of the spatial dependence of the effective mass}

A necessary condition for the applicability of the one-band
equations for the envelope functions [one equation is understood
here, valid near the bottom of the non-degenerate conduction band, or a system of
equations for the degenerate valence band] used in the effective-mass method is
``shallowness'' of these states: their energy, measured from the
band edge, should be small in comparison with the inter-band
energy. Therefore, bearing in mind the one-band version of
the effective-mass method, we restrict the discussion to heterostructures
consisting of related materials, where the band
offsets are small in comparison with the characteristic
band gaps; this means, as a rule, that other band parameters of the
semiconductors differ only slightly.
Let us consider the first problem, which arises even for heterostructures
whose chemical composition varies smoothly on scales of the order of $a$.
As the zeroth-order potential we
choose the potential of the crystalline lattice, continued to all
space, of one of the materials of the structure [this is not a
unique choice, see Ref. \cite{smith}], and we treat the difference between
the potentials of the lattices of the remaining semiconductors
as a small perturbation. Following the approach of Luttinger and Kohn and deriving the
many-band ${\bf k\cdot p}$ system of equations (see, e.g., Ref. \cite{leibler}), we
can next attempt to solve the problem of the correct order of the non-commuting operators
in the kinetic-energy operator for the one-band equations.
But here yet another problem arises.

Reduction of the many-band system of equations to a
one-band effective-mass equation is achieved by eliminating
the small envelope functions from the many-band ${\bf k\cdot p}$
system in favor of the large ones by means of some procedure.
We make a small departure here and make use of a formal
analogy between the relativistic Dirac equation and the
many-band ${\bf k\cdot p}$ system of equations for the envelope
functions \cite{keldysh}, which is most simply seen in the two-band
approximation (the conduction band and the non-degenerate
valence band). In the relativistic theory there are two approaches
to deriving an equation for shallow electron states.
One of them consists in eliminating the small positron
component of the wave function by the method of substitution. In
this case, we obtain either an exact equation for the electron
component, which is not an eigenvalue equation, Ref.~\cite{messia}, Ch.~20, Sec.~28,
or an approximate equation whose Hermitian nature must be checked separately
\cite{landavshits}. The second approach is a Foldy-Wouthuysen transformation, an
approximate unitary transformation of the Dirac equation, Ref.~\cite{messia},
Ch.~20, Sec.~33.

In our case the first approach is comparatively simple to
realize only in the two-band approximation, see, e.g., Ref. \cite{suris}.
In a treatment of the contribution of distant bands, and
this is necessary, in particular, for a valid description of the
heavy holes, a number of problems arises.
Thus, the authors of Refs. \cite{burt} and \cite{foreman2} were able to take into
account only a few of the first-order corrections to the ``standard''
Kohn-Luttinger equation with position-independent
effective mass [the small parameter here is the ratio of the
characteristic band offset to the characteristic inter-band
energy]. However, treating the expression obtained, for
example, in Ref. \cite{burt} for the position-dependent effective
mass, it can be shown that the effective mass of the edge of
the conduction band of one of the non-basis semiconductors
does not contain inter-band matrix elements of the perturbation potential,
obtained using Bloch functions of the band edge of the basis crystal
(see Sec. 4.1 below). It can be easily
seen that this is equivalent to the poorly justified approach of
neglecting the difference between the inter-band matrix elements of
the momentum operator or, what is equivalent, the
difference between the Bloch functions for the materials
making up the structure.

Hence it follows that we should give special attention to
the problem of taking distant bands into account. Efforts at
solving it by direct elimination of the small envelope
functions by the method of substitution, in addition to its laboriousness,
lead finally to a non-Hermitian equation, whose solution is still in need of
a valid interpretation.

Below we will follow the second approach, i.e., we will
apply the unitary transformation eliminating the small envelope
functions \cite{kohn-lut,bir-pikus}. Since we are considering heterostructures
consisting of related materials, the standard effective-mass
method will play the role of a first approximation. An account of
the spatial dependence of the effective mass necessitates
treating corrections to the standard theory, where it is
necessary to take into account all corrections of the same
order without the accuracy excess.

In order to understand what corrections should be taken
into account, let us turn to the relativistic analogy with the
hypothetical Dirac equation containing the inhomogeneous
gap $2m ( {\bf r} )c^2$, where $c$ is the speed of light in vacuum (see
Appendix A). The ordinary one-band effective-mass equation
is an analog of the non-relativistic Schr\"odinger equation.
It is important, however, that the effective mass in the two-band
approximation is proportional to the local band gap $E_g ( {\bf r} )$
[this is valid if the effective mass is formed mainly by
${\bf k\cdot p}$ interaction], and its relative variation
${\delta m}/m\simeq {\delta E_g}/E_g$. Since the correction to the kinetic energy
describing the spatial dependence of the effective mass will
have a ``relativistic'' character, the desired equations for the
heterostructures will be analogous to the Schr\"odinger equation
with all relativistic corrections---both the usual ones (the
contribution of nonparabolicity of the dispersion law, proportional to
${\bf p}^4$, where ${\bf p}$ is the momentum operator; the contribution
of the spin-orbit interaction; and the so-called Darwin term,
proportional to the second derivative of the
potential energy) and a new pseudo-relativistic correction
describing $\delta m ( {\bf r} )$. Of course, the present arguments are valid
for describing states whose energies, reckoned from the band
edge of any of the materials making up the structure, are of
the order of the band offset. The case of a very small
band offset, where the offset is small in comparison
with the energies of the states is quite trivial:
depending on the energies of the states under consideration an
account of the spatial dependence of the effective mass can
require treating terms with higher and higher powers of the
momentum operator. We will not consider such a situation.
In this sense, introducing a term proportional to the fourth
power of the momentum operator into the effective-mass
equation is a necessary condition for a consideration of the
effective mass spatial dependence. Note that for homogeneous
semiconductors an effective-mass equation analogous to the Schr\"odinger
equation with first relativistic corrections was discussed
already in Ref. \cite{bir-pikus} (Sec. 27).

A typical shortcoming of previous works dedicated to a
generalization of the effective-mass method to electron states
in heterostructures is that they take account within the
framework of perturbation theory of only some of the terms of
a given order. Thus, Refs. \cite{leibler}-\cite{kisin} take account of the spatial
dependence of the effective mass parameters, but neglect
corrections $\propto {\bf p}^4$. In Refs. \cite{leibler, smith, young, burt},
dedicated to deriving the one-band equations for the envelope functions
directly from a many-band ${\bf k\cdot p}$ system, the main error,
leading to an invalid result, is an incorrect estimate, according to
which the contribution of the ${\bf k\cdot p}$ interaction terms [i.e.,
the terms $\hbar {\bf k}{\bf p}_{nn'}/m_0$, where $m_0$ is the free electron mass and
${\bf p}_{nn'}$ is the inter-band matrix element of the momentum] is of
order the contribution of the potential energy terms [by potential energy
here we mean the difference in the periodic
potentials of the semiconductors making up the structure,
also treated in the perturbation theory context]. In the case of a
smooth heterojunction the correct procedure for deriving the
one-band effective-mass equation near the bottom of the
conduction band with all the above-indicated contributions
taken into account was followed in Ref. \cite{we1}.

\subsection{Account of atomically sharp heterojunctions}

The second problem requiring careful study is the
non-smooth nature of actual heterojunctions, where the transition
from one material to the other occurs over scales of the order
of $a$. In this case, first, Leibler's many-band ${\bf k\cdot p}$
system \cite{leibler}, where smoothness of the potential was a necessary condition
for validity of the system, is in need of refinement, and
second, the problem of transforming to ${\bf r}$ space from the region
of ${\bf k}$ space bounded by the first Brillouin zone is more
complicated \cite{kohn-lut}. It is also necessary to analyze the
consequences of the unitary transformation eliminating distant
bands. It is important to estimate the error that enters at each
step. An estimate of this error either gives us confidence in
the absence of an excess of accuracy or it challenges the
validity of the effective-mass approximation. In the works
known to us which treat sharp heterojunctions, such an estimate
is lacking. References \cite{burt} and \cite{foreman2}, for example,
only point to its smallness, and Ref. \cite{foreman1} made some
approximations whose accuracy were not even estimated.

Thus, we can formulate the following steps in the
construction of an effective-mass approximation for
heterostructures: a) obtaining a many-band ${\bf k\cdot p}$
system of equations for the envelope functions taking proper account of possible
sharpness of the heterojunction; b) reducing this system to
one-band equations with the help of the unitary transformation to
${\bf k}$ space, transforming to ${\bf r}$ space and transforming the
resulting equation to differential form; c) estimating the accuracy of
these transformations. Following this scheme, step a) is realized in Sec. 2.
The equations include contributions associated with the non-smoothness of
the heterojunction on scales of order $a$ which are treated within the framework of
an approach similar to that used in Ref. \cite{bir-pikus} to describe the
short-range part of the impurity potential. Section 3 shows
that a common differential equation over all space for the
envelope functions for sharp heterojunctions exists, and its
accuracy is determined by the procedure of transforming to
one-band equations in ${\bf r}$ space. One-band equations are considered in
Sec. 4.1 (the conduction band)  and in Sec. 4.2 (the
valence band). Section 4.3 is devoted to heterostructures
with super-thin layers. It is shown that additional contributions to the ``standard''
effective-mass equation can be classified by powers of the small parameter
$\bar k a_m$, where $\bar k$ is a characteristic value of the quasi-momentum of the
state and $a_m$ is of the order of the lattice constant. Section 5 constructs
a hierarchical scheme of effective-mass equations, the $n$th
level of which corresponds to taking account of these additional
contributions up to $(\bar k a_m)^n$ inclusively. The zeroth order of the
hierarchy ($n=0$) corresponds to the ``standard'' effective-mass
equation with position-independent parameters. At the
first level of the hierarchy each heteroboundary gives an additional
$\delta$-function contribution to the potential energy. Only
at the second level of the hierarchy does spatial dependence
of the effective mass appear, along with corrections associated
with weak nonparabolicity of the spectrum and heterointerface
terms of the spin-orbit interaction. At higher levels
of the hierarchy nonlocal contributions arise, and the one-band
differential effective-mass equations do not exist. Results are compared
with the conclusions of other authors. Brief reports on the results
obtained have been published elsewhere \cite{we2}-\cite{we4}.

\section{Many-band ${\bf k\cdot p}$ system of equations for envelope functions
in the case of a sharp (001) heterojunction}

Let us consider a heterojunction formed from related
lattice-matched semiconductors with zinc-blende structure.
The Schr\"odinger equation without relativistic corrections,
which will be taken into account below, and in the absence
of external potentials has the usual form
\[
\left( \frac{{\bf p}^2}{2m_0}+ U ( {\bf r} )\right)
\Psi \left( {\bf r}\right) =\epsilon
\Psi \left( {\bf r}\right) .
\]
Here $U ( {\bf r} )\equiv U$ is the crystal potential of the heterostructure.
To start with, we will use the following model of this potential:
\begin{equation}
U=U_1+G \left( z\right)\left[ U_2-U_1\right]
\equiv U_1+G\left( z\right) \delta U,
\label{lattice}
\end{equation}
where $U_1\equiv U_1 ( {\bf r} )$ and $U_2\equiv U_2 ( {\bf r} )$
are periodic (with the same period) potentials, continued through the entire
structure, of the left-hand and right-hand materials, respectively,
the $z$ axis is directed perpendicular to the plane of the heterojunction,
$G(z)$ is the form factor of the heterojunction
\[
G (z)\bigm|_{z<-d} =0, \quad G (z)\bigm|_{z>d} =1;
\]
and the width of the transitional region of the heterojunction
is $2d$ [non-one-di\-men\-siona\-lity of $G(z)$ will be taken into account below].

It is natural to treat the potential $G( z) \delta U$ as a perturbation.
As the basis for expanding the wave function we use
the complete orthonormal set of Kohn-Luttinger functions
$\{u_{n0}e^{ i{\bf k\cdot r} }\}$:
\begin{equation}
\Psi \left( {\bf r}\right) =\sum_{n^{\prime }}\int {\cal F} _{n^{\prime }}
\left( {\bf k}^{\prime }\right) e^ { i{\bf k}^{\prime }
{\bf \cdot r}} u_{n^{\prime }0}\, d^3k^{\prime },  \label{expans}
\end{equation}
where $u_{n0}\equiv u_{n0} ({\bf r} ) $ is the periodic Bloch amplitude for the
edge $\epsilon_{n0}$ of the $n$th band of the left-hand crystal at the $\rm \Gamma $
point of the Brillouin zone (in the non-relativistic limit),
\[
\left( \frac{{\bf p}^2}{2m_0}+U_1\right) u_{n0}=\epsilon _{n0}u_{n0}.
\]
óThe sum in Eq. (\ref{expans}) is over all bands, and the integral
is over the Brillouin zone here and in what follows unless
otherwise stated; ${\cal F}_n ( {\bf k} ) $ is the envelope function for the
$n$ band in ${\bf k}$ space. Following the standard procedure \cite{kohn-lut},
we obtain a system of ${\bf k\cdot p}$ equations \cite{leibler}:
\[
\left( \epsilon _{n0}+\frac{{\hbar}^2{\bf k}^2}{2m_0}\right) {\cal F}_n\left( {\bf k}
\right) +\sum_{n^{\prime }}\frac{{\hbar} {\bf p}_{nn^{\prime }}\cdot 
{\bf k}}{m_0}{\cal F}_{n^{\prime }}\left( {\bf k}\right) + 
\]
\begin{equation}
+\sum_{n^{\prime }}\int {\cal M}_{nn^{\prime }}
\left( {\bf k},{\bf k}^{\prime }\right) {\cal F}_{n^{\prime }}
\left( {\bf k}^{\prime }\right) \, d^3 k^{\prime }
=\epsilon {\cal F}_n\left( {\bf k}\right) ;  \label{kp}
\end{equation}
\[
{\cal M}_{nn^{\prime }}\left( {\bf k},{\bf k}^{\prime }\right) =\sum_{j
}C_j^{nn^{\prime }}{\cal G}\left( k_z-k_z^{\prime }+K_{zj}\right)
\delta \left( {\bf k}_{||}-{\bf k}_{||}^{\prime }+{\bf K}_{||j}\right) .
\]
Here ${\bf p}_{nn^{\prime }}=\left\langle n\mid {\bf p}\mid n^{\prime
}\right\rangle $ and $C_j^{nn^{\prime }}=\left\langle n\mid \delta U
e^{ i{\bf K}_j{\bf r} } \mid n^{\prime }\right\rangle $, and the matrix
elements of a periodic operator ${\bf f}$ are defined as follows:
\[
\left\langle n\mid {\bf f}\mid n^{\prime }\right\rangle
=\frac{\left( 2\pi \right) ^3}\Omega
\int_{\rm cell}u_{n0}^*{\bf f}u_{n^{\prime }0}\, d^3r,
\]
where $\Omega $ is the volume of the unit cell; ${\bf k}_{||} = ( k_x,k_y,0 ) $,
and $K_{zj}$ and ${\bf K}_{||j}$ are the components of the vector ${\bf K}_j $
of the inverse lattice perpendicular and parallel to the plane of the
heteroboundary, respectively; ${\cal G} ( k_z ) $ is the Fourier transform
of $G ( z )$. Let us analyze the expression for the matrix elements
${\cal M}_{nn^{\prime }} ( {\bf k},{\bf k}^{\prime} ) $:
\[
{\cal M}_{nn^{\prime }}\left( {\bf k},{\bf k}^{\prime }\right) =\sum_{%
j\ ({\bf K}_{||j}=0)}C_j^{nn^{\prime }}{\cal G}\left(
k_z-k_z^{\prime }+K_{zj}\right) \delta \left( {\bf k}_{||}-{\bf k}%
_{||}^{\prime }\right) + 
\]
\begin{equation}
+\sum_{j\ ({\bf K}_{||j}\neq 0)}C_j^{nn^{\prime }}{\cal G}\left(
k_z-k_z^{\prime }+K_{zj}\right) \delta \left( {\bf k}_{||}-{\bf k}%
_{||}^{\prime }+{\bf K}_{||j}\right) .  \label{sums}
\end{equation}
The second sum in Eq. (\ref{sums}) describes transfer processes in the
two-dimensional Brillouin zone, when the projections of any pair of vectors
${\bf k}$ and ${\bf k}^{\prime}$ from the bulk Brillouin zone onto
the plane of the heterojunction satisfy the condition
${\bf k}_{||}^{\prime}-{\bf k}_{||}={\bf K}_{||j}\neq 0$.
For a heterojunction of arbitrary orientation such transfer processes exist.
However, for the orientation of interest to us---the (001) orientation---their
contribution to the desired equations for the envelope functions disappears (see
Appendix B):
\begin{equation}
{\cal M}_{nn^{\prime }}\left( {\bf k},{\bf k}^{\prime }\right) = 
\delta \left( {\bf k}_{||}-{\bf k}_{||}^{\prime }\right) \left[ {\cal G}
\left( k_z-k_z^{\prime }\right) \delta U_{nn^{\prime }}+\sum_{j\neq 0}
C_j^{nn^{\prime }}{\cal G}\left( k_z-k_z^{\prime }+K_j\right)
\right] ,  \label{int}
\end{equation}
where we have introduced the notation: $K_j = ( 4\pi /a )j$; $j=\pm 1, \pm 2, ...$;
and $\delta U_{nn^{\prime }}=C_0^{nn^{\prime }}$. If $G ( z ) $ is a sufficiently
smooth function, $a \ll d$, and we are interested in states with
$\bar{k}_z \ll 2\pi /a$, where $\bar{k}_z$ is a characteristic value of the
quasi-momentum of the state, we can neglect the second term inside
the brackets in Eq. (\ref{int}) and as a result obtain the well-known
set of equations for the envelope functions \cite {leibler,we1}.
In the case of an atomically sharp heterojunction, on the other hand, it is
possible to proceed in the spirit of the method used in Ref. \cite{bir-pikus}
to describe a short-range impurity potential. We introduce
the function $G^{\prime }(z)\equiv dG(z) /dz$, localized on the heteroboundary,
$|z| \leq d$. Then for $j \neq 0$ we have:
\[
{\cal G}\left( k_z-k_z^{\prime }+K_j\right) =\frac 1{2\pi
}\int\limits_{-\infty }^{+\infty }G\left( z\right) e^ { -i\left(
k_z-k_z^{\prime }+K_j\right) z } \, dz= 
\]
\[
=\frac 1{2\pi i}\frac 1{k_z-k_z^{\prime }+K_j}\int\limits_{-\infty
}^{+\infty }G^{\prime }(z) e^{ -i\left( k_z-k_z^{\prime }+K_j\right) z }\,
dz=
\]
\begin{equation}
=\frac 1{2\pi iK_j}\left( 1-\frac{k_z-k_z^{\prime }}{K_j}+
\ldots \right) \int\limits_{-d}^d G^{\prime }(z) e^{
-iK_jz } \left[ 1-i\left( k_z-k_z^{\prime }\right) z+
\ldots \right]\, dz,  \label{chems}
\end{equation}
and we can write the sum in Eq. (\ref{int}) in the form of an expansion
in powers of $(k_z-k_z^{\prime })$:
\begin{equation}
\sum_{j\neq 0}C_j^{nn^{\prime }}{\cal G}\left( k_z-k_z^{\prime
}+K_j\right) =\sum_{s=0,1,2,\ldots}\frac {(k_z-k_z^{\prime })^s}{2\pi }
D_{snn^{\prime }}.  \label{dterms}
\end{equation}
The constants in the expansion (\ref{dterms}) have the form
\[
D_{0nn^{\prime }}=\sum_{j\neq 0}C_j^{nn^{\prime }}\frac 1{iK_j}
\int\limits_{-d}^dG^{\prime }(z) e^{ -iK_jz }\, dz,
\]
\[
D_{1nn^{\prime }}=\sum_{j\neq 0}C_j^{nn^{\prime }}\frac 1{iK_j}
\int\limits_{-d}^dG^{\prime }(z) e^{ -iK_jz } \left(
-\frac 1{K_j}-iz \right)\, dz, \ldots
\]
The present approach fundamentally allows one to treat even
mathematically sharp heterojunctions, since the necessary
convergence of the coefficients $D_{lnn^{\prime }}$ is ensured by the
property $C_j^{nn^{\prime }}\rightarrow 0$ as $K_j \rightarrow \infty $
[but a physically realizable heterojunction cannot be mathematically
sharp].

Let us consider the corrections associated with sharpness
of the heterojunction. Simple estimates show that terms proportional to
$D_{0nn^{\prime }}$, $D_{1nn^{\prime }}$,~\dots can give corrections not
greater in order of magnitude than $a \bar{k}_z$, $(a \bar{k}_z) ^2$,~\dots,
respectively. Our goal is to obtain one-band equations with
position-dependent effective-mass parameters, which is
achieved by taking account of corrections of order $( \lambda \bar{k}_z) ^2$
to the standard approximation. Here we have introduced a characteristic
``two-band'' length $\lambda =\hbar (2mE_g) ^{-1/2}$.
For GaAs, for example, $\lambda \approx 6${\rm \AA }. We will make use of the smallness of
the parameter $d \cdot \bar k_z$, which will allow us to write down the
final equation in quite simple form (see below). Thus, three
quantities having the dimensions of length, $a$, $d$ and $\lambda$, in
combination with $\bar k_z$ form three parameters whose smallness
is employed in the present method. In our view, the situation
$a$~{\amsfnt\char"2E}~$d$~{\amsfnt\char"2E}~$\lambda$, is the most realistic, being realized in
semiconductor heterostructures with sharp heteroboundaries. Thus, the
parameter $\lambda \bar k_z$ can be taken as the main small parameter of
the problem and the sum in Eq. (\ref{dterms}) can be restricted to terms
with $s=0$ and $s=1$.

As a result, the many-band system of ${\bf k\cdot p}$ equations (\ref{kp}) takes the form
\[
0=\left( \epsilon _{n0}-\epsilon +\frac{ {\hbar}^2 {\bf k}^2 }{2m_0}\right)
{\cal F}_n\left( {\bf k}\right) +\sum_{n^{\prime }}\frac{{\hbar}{\bf p}%
_{nn^{\prime }}\cdot {\bf k}}{m_0}{\cal F}_{n^{\prime }}\left( {\bf k}%
\right) + 
\]
\begin{equation}
+\sum_{n^{\prime }}\int \left[ {\cal G}\left( k_z-k_z^{\prime
}\right) \delta U_{nn^{\prime }}+\frac 1 {2\pi}D_{0nn^{\prime }}
+\frac {k_z-k_z^{\prime }}{2\pi}D_{1nn^{\prime }} \right]
{\cal F}_{n^{\prime }}\left( k_z^{\prime },{\bf k}%
_{||}\right) \, dk_z^{\prime }.  \label{kpa}
\end{equation}
Here we can distinguish different contributions of the perturbation potential:
the contribution of the ``smooth'' part is represented by the first term inside the brackets,
and the contribution of the ``sharp'' part is represented by the second
and third terms.

\subsection{Account of the 3D character of the form factor}

Let us now consider instead of Eq. (\ref{lattice}) a more realistic
form of the heteropotential:
\begin{equation}
U=U_1+g \left( {\bf r}_{||},z\right) \delta U, \label{real lattice}
\end{equation}
where ${\bf r}_{||}=(x,y,0)$. By definition,
$g ({\bf r}_{||},z)\bigm|_{z<-d} =0$ and $g ({\bf r}_{||},z)\bigm|_{z>d} =1$,
and the function $g({\bf r}_{||},z)$ is periodic in ${\bf r}_{||}$. For our case of (001)
heterostructure the unit translation vectors in the plane of the heteropotential are
${\bf a}_1=(1,1,0)a/2$ É ${\bf a}_2=(-1,1,0)a/2$. The sites of the two-
dimensional inverse lattice with basis vectors ${\bf b}_1=(1,1,0)2\pi/a$ and
${\bf b}_2=(-1,1,0)2\pi/a$ are projections of the sites of the three-dimensional
inverse lattice onto the (001) plane.

We expand $g({\bf r}_{||},z)$ in a 2D Fourier series:
\[
g\left( {\bf r}_{||},z \right)=\sum_{l}G_l \left( z \right)
e^{ i{\bf K}_l {\bf r}_{||} },
\]
where the summation index $l$ is defined so that the vectors ${\bf K}_l$
determine the sites of the indicated two-dimensional inverse lattice, and
\[
G_l \left( z \right)
=\frac{1}{\Omega_{||}} \int_{\rm 2D \atop cell} g\left( {\bf r}_{||},z\right)
e^{ -i{\bf K}_l {\bf r}_{||} } \, {d}^2r_{||};
\]
The integration is over a unit cell of the 2D lattice with area
$\Omega_{||}=a^2/2$. Denoting $G(z)\equiv G_0(z)$, we obtain for the
perturbation potential
\begin{equation}
g \left( {\bf r}_{||},z\right) \delta U =G\left( z\right) \delta U +
\delta U \sum_{l\neq 0} G_l \left( z \right)
e^{ i{\bf K}_l {\bf r}_{||} } . \label{null plus}
\end{equation}
It can be seen that the simple model (\ref{lattice}) takes into account
the first term in the sum (\ref{null plus}). The functions $G_l ( z )$ for $l \neq 0$
are nonzero only for $z \in [-d,d]$. Therefore, the left-hand side of Eq. (\ref{kp})
will include an additional sum of interface contributions:
\[
\sum_{n^{\prime }}\int {\cal M}_{nn^{\prime
}}^{||}\left( {\bf k},{\bf k}^{\prime }\right) {\cal F}_{n^{\prime }}
\left( {\bf k}^{\prime }\right)\, {d}^3 k^{\prime },
\]
where
\[
{\cal M}_{nn^{\prime }}^{||}\left( {\bf k},{\bf k}^{\prime }\right)
=\sum_{l \neq 0;j}\frac{C_j^{nn^{\prime }}}{\left( 2\pi \right)^3}
\int_{\rm all \atop space} G_l \left( z \right)
e^{ i{\bf K}_l {\bf r}_{||} }
e^{ -i \left( {\bf k}-{\bf k}^{\prime }+{\bf K}_j\right){\bf r}
} \, {d}^3 r.
\]

For states with $\left| k_x\right| +\left| k_y\right|<\pi /a$
(see Appendix B) we obtain
\[
{\cal M}_{nn^{\prime }}^{||}\left( {\bf k},{\bf k}^{\prime }\right)
=\delta \left( {\bf k}_{||}-{\bf k}_{||}^{\prime} \right)
\sum_{l \neq 0;j}\frac{C_j^{nn^{\prime }}}{2\pi}
\delta_{ {\bf K}_l,{\bf K}_{||j} }
\int\limits_{-d}^d G_l \left( z \right)
e^{ -i \left( k_z-k_z^{\prime }+ K_{zj}\right)z }\,  {d}z=
\]
\[
=\delta \left( {\bf k}_{||}-{\bf k}_{||}^{\prime} \right)
\sum_{j ({\bf K}_{||j} \neq 0)}\frac{C_j^{nn^{\prime }}}{2\pi \Omega_{||}}
\int\limits_{-d}^d \, {d}z \int_{\rm 2D \atop cell}\,
{d}^2r_{||} g\left( {\bf r}_{||},z\right)
e^{ -i{\bf K}_{||j} {\bf r}_{||} }
e^{ -i \left( k_z-k_z^{\prime }+ K_{zj}\right)z }.
\]
For a smooth heterojunction it is necessary to retain only
terms with $K_{zj}=0$ in the sum and develop the expression in
the standard way \cite {leibler}, since the equations for the envelope
functions in the ${\bf r}$ representation will include an additional
interface potential. Note that while the smooth part of $G(z)\delta U$
ensures mixing of states of the same crystal symmetry [i.e., the
local symmetry of the smooth part of the perturbation potential
$G(z)\delta U$ coincides with the symmetry of the bulk crystal],
this additional potential also ensures mixing of states of different symmetry.
We are interested in the case $d \cdot \bar{k}_z \ll 1$, and this obviates the
necessity of separating effects of this additional potential into contributions
of smooth and sharp parts and allows us to use the expansion
\begin{equation}
{\cal M}_{nn^{\prime }}^{||}\left( {\bf k},{\bf k}^{\prime }\right)
=\delta \left( {\bf k}_{||}-{\bf k}_{||}^{\prime} \right)
\sum_{s=0,1,2,\ldots}\frac {(k_z-k_z^{\prime })^s}{2\pi}
D_{snn^{\prime }}^{||} , \label{dterms plus}
\end{equation}
where
\[
D_{0nn^{\prime }}^{||}= \sum_{j ({\bf K}_{||j} \neq 0)}
\frac{C_j^{nn^{\prime }}}{\Omega_{||}} \int\limits_{-d}^d\, {d}z
\int_{\rm 2D \atop cell} \, {d}^2r_{||} g\left( {\bf r}_{||},z\right)
e^{ -i{\bf K}_j {\bf r} },
\]
\[
D_{1nn^{\prime }}^{||}= \sum_{j ({\bf K}_{||j} \neq 0)}
\frac{C_j^{nn^{\prime }}}{\Omega_{||}} \int\limits_{-d}^d \, {d}z
\int_{\rm 2D \atop cell} \, {d}^2r_{||} g\left( {\bf r}_{||},z\right)
e^{ -i{\bf K}_j {\bf r} } \left( -iz \right), \ldots
\]
In the expansion (\ref{dterms plus}) it is necessary to keep only the first
two terms; the terms proportional to $D_{0nn^{\prime }}^{||}$ É
$D_{1nn^{\prime }}^{||}$ can give contributions of order $d \cdot \bar k_z$ and
$(d \cdot \bar k_z )^2$, respectively.

We have shown that taking the three-dimensionality of
the form factor into account (see Eq. (\ref{real lattice})), causes no special
difficulty for analysis, and we now make an important observation
which will allow us to use the simple model (\ref{lattice}).
The function $g({\bf r}_{||},z)$ has lower symmetry than $G(z)$.
Specifically, it is invariant under symmetry transformations from
the point group $\rm C_{2v}$. But the complete perturbation potential in
both models, (\ref{lattice}) and (\ref{real lattice}), possesses the
same symmetry, both its point-group symmetry ($\rm C_{2v}$) and translational
symmetry in the plane of the heterojunction. Information about
$\rm C_{2v}$ symmetry will be preserved, however, only if the contribution
of the sharp part of the potential $G(z) \delta U$ in model (\ref{lattice}) is
taken into account. Therefore, using model (\ref{real lattice}) instead of model
(\ref{lattice}) does not give anything qualitatively new, and only leads to renormalization
of some parameters, namely those that are negligibly small
for the smooth heterojunction in model (\ref{lattice}). The expressions
for these parameters are very complicated, so in what follows we will stick
with model (\ref{lattice}).

\subsection{Account of relativistic corrections}

Let us now consider relativistic effects. We restrict the
discussion to the spin-orbit interaction. The remaining
relativistic contributions only influence the values of the constants
that we will obtain. We assume that within the framework of perturbation
theory the characteristic parameter of the spin-orbit interaction and
also the difference of this parameter for the left-hand and right-hand crystals are less
than or of the same order as the characteristic band offset.
The expansion of the total wave function, as before,
is given by expression (\ref{expans}). Omitting intermediate manipulations,
we give the resulting, quite lengthy ${\bf k\cdot p}$ system of
equations which take into account the spin-orbit interaction
\[
\left( \epsilon _{n0} -\epsilon +\frac{{\hbar }^2{\bf k}^2}{2m_0} \right)
{\cal F}_n\left( {\bf k}\right) +\sum_{n^{\prime }}\frac{\hbar {\bf p}
_{nn^{\prime
}}\cdot {\bf k}}{m_0}{\cal F}_{n^{\prime }}\left( {\bf k}\right) +
\]
\[
+\sum_{n^{\prime }}\delta U_{nn^{\prime }}\int {\cal G}\left(
k_z-k_z^{\prime }\right) {\cal F}_{n^{\prime }}\left( k_z^{\prime },%
{\bf k}_{||}\right) \, {d}k_z^{\prime }
+\sum_{n^{\prime }}\frac{\hbar \left\langle n\right| \left[ \nbla %
U_1\times {\bf p}\right] \left| n^{\prime }\right\rangle \cdot \sgma }
{4m_0^2c^2}{\cal F}_{n^{\prime }}\left( {\bf k}\right) +
\]
\[
+\sum_{n^{\prime }}\frac{\hbar \left\langle n\right| \left[ \nbla \delta
U\times {\bf p}\right] \left| n^{\prime }\right\rangle \cdot \sgma }
{4m_0^2c^2}\int {\cal G}\left( k_z-k_z^{\prime }\right) {\cal
F}_{n^{\prime }}\left( k_z^{\prime },{\bf k}_{||}\right)\, {d}k_z^{\prime }+
\]
\[
+\sum_{n^{\prime }}\frac{\hbar \left\langle n\right| \left[ {\bf n} \delta
U\times {\bf p}\right] \left| n^{\prime }\right\rangle \cdot \sgma }
{4m_0^2c^2}\int i\left( k_z-k_z^{\prime }\right)
{\cal G}\left( k_z-k_z^{\prime }\right) {\cal F}_{n^{\prime }}
\left( k_z^{\prime },{\bf k}_{||}\right) \, {d}k_z^{\prime }+
\]
\[
+\sum_{n^{\prime }}\int \frac{{\hbar}^2 \left\langle n\right| \left[ \nbla
\delta U\times {\bf k}^{\prime}\right] \left| n^{\prime }\right\rangle
\cdot \sgma } {4m_0^2c^2} \delta \left( {\bf k}_{||}-{\bf k}_{||}^{\prime }
\right) {\cal G}\left( k_z-k_z^{\prime }\right) {\cal F}
_{n^{\prime }} \left( {\bf k}^{\prime } \right) \, {d}^3 k^{\prime }+
\]
\[
+\sum_{n^{\prime }}\int \left( \frac 1{2\pi}D_{0nn^{\prime }}
+\frac {k_z-k_z^{\prime}}{2\pi}D_{1nn^{\prime }}\right){\cal F}
_{n^{\prime }}\left( k_z^{\prime },{\bf k}_{||}\right)\, {d}k_z^{\prime }+
\]
\[
+\sum_{n^{\prime }}\int \left( \frac 1{2\pi}{\bf S}_{0nn^{\prime }}+ \frac
{k_z-k_z^{\prime }}{2\pi}{\bf S}_{1nn^{\prime }} \right) \cdot \sgma {\cal F}
_{n^{\prime }}\left( k_z^{\prime },{\bf k}_{||}\right) \, {d}k_z^{\prime }+
\]
\begin{equation}
+\sum_{n^{\prime }}\int \frac 1{2\pi}
\delta \left( {\bf k}_{||}-{\bf k}_{||}^{\prime } \right)
\left[ \hbar{\bf k}^{\prime} \times \sgma \right] \cdot
{\bf B}_{0nn^{\prime }} {\cal F}_{n^{\prime }}\left( {\bf k}^{\prime }
\right) \, {d}^3 k^{\prime} =0. \label{all band kp}
\end{equation}
The vectors ${\bf S}_{0nn^{\prime }}$, ${\bf S}_{1nn^{\prime }}$ and
${\bf B}_{0nn^{\prime }}$ have the following form:
\[
{\bf S}_{0nn^{\prime }}=
\sum_{j\neq 0}\frac{\hbar \left\langle n\right| \left[ {\nbla }
\left( e^{ iK_jz } \delta U\right) \times {\bf p}%
\right] \left| n^{\prime }\right\rangle }{4 iK_jm_0^2c^2}
\int\limits_{-d}^dG^{\prime }(z) e^{ -iK_jz } \, {d}z;
\]
\[
{\bf S}_{1nn^{\prime }}=-\sum_{j\neq 0}\frac{\hbar \left\langle
n\right| \left[ {\nbla }(e^{ iK_jz } \delta U)\times {\bf %
p}\right] \left| n^{\prime }\right\rangle }{4 K_jm_0^2c^2}
\int\limits_{-d}^dG^{\prime }(z) e^{ -iK_jz} z\, {d}z- 
\]
\[
-\sum_{j\neq 0}\frac{\hbar \left\langle n\right| e^{ iK_jz }
\left[ \nbla \delta U\times {\bf p}\right] \left| n^{\prime
}\right\rangle }{4i K_j^2m_0^2c^2}\int\limits_{-d}^dG^{\prime }(z)
e^{ -iK_jz } \, {d}z;
\]
\[
{\bf B}_{0nn^{\prime }}=
\sum_{j\neq 0}\frac{ \hbar \left\langle n\right| \nbla
\left( e^{ iK_jz } \delta U\right) \left| n^{\prime }
\right\rangle }{4 iK_jm_0^2c^2}
\int\limits_{-d}^dG^{\prime }(z)e^{ -iK_jz }\, {d}z.
\]
Here ${\bf n}$ is the unit vector along the $z$ axis, ${\bf n}G^{\prime }
( z) \equiv \nbla G( z) $, and $\sgma$ are the Pauli matrices. On the left-hand side of
Eq. (\ref{all band kp}) the fourth term describes the spin-orbit interaction
in the potential of the basis semiconductor; the fifth,
sixth, and seventh terms are due to the smooth part of the
perturbation potential. The terms proportional to
${\bf S}_{0nn^{\prime }}$, ${\bf S}_{1nn^{\prime }}$ and
${\bf B}_{0nn^{\prime }}$ are due to the sharpness of the potential. In Ref. \cite{we1},
in a consideration of the state of the conduction
band in heterostructures with smooth heteroboundaries, we
neglected the sixth and seventh terms on the left-hand side of
Eq. (\ref{all band kp}) as small.  We noted that in second-order perturbation
theory they, together with $\hbar {\bf k\cdot p}_{nn^{\prime }}/m_0$, give a correction
only of order $( \lambda \bar{k}_z) ^2m/{m _0}$, which can be neglected as the
effective mass is small in comparison with $m _0$. For the hole
states, on the other hand, $m/m _0$ is able not to be a small
parameter.

We do not consider ${\bf k}$-linear contributions of the spin-orbit
interaction due to the potential $U_1$. They give corrections of order
$( \lambda \bar{k}_z) ^3$ (third-order corrections, along with two
terms of the form× $\hbar {\bf k\cdot p}_{nn^{\prime }}/m_0$) similar to the contribution
responsible for removing the spin degeneracy in the conduction band of the bulk
semiconductor (we neglect terms of this order), and for the valence band it is well
known that to first order the contribution of these terms is small, and to
second order, along with $\hbar {\bf k\cdot p}_{nn^{\prime }}/m_0$ they only renormalize
the effective-mass parameters.

As for the contributions from the sharpness of the heterojunction
potential to the spin-orbit interaction, the terms
proportional to ${\bf S}_{0nn^{\prime }}$, can give corrections of order
$a \bar{k}_z$, while the terms ${\bf S}_{1nn^{\prime }}$ and
${\bf B}_{0nn^{\prime }}$ can give corrections of order $(a \bar{k}_z) ^2$.

It is trivial to generalize to the case of many heterojunctions.
In this case it is convenient to choose the coordinates of the
heteroboundaries so that the distances between them be integer multiples of
$a/2$, so that the phase factor of each of the expansions of the type
(\ref{dterms}) is equal to unity.

\section{Problem of transformation of the effective-mass equations from quasi-momentum space
to coordinate space}

In the following section we obtain the one-band
effective-mass equations for the conduction band and the valence band.
But first of all, we must discuss a problem arising in the method of envelope functions and associated with
the boundedness of ${\bf k}$ space. Let us consider the following
one-band equation for the envelope functions $f( k_z )$ in ${\bf k}$ space:
\begin{equation}
\int H\left( k_z,k_z^{\prime }\right) f\left( k_z^{\prime }\right)\, 
{d}k_z^{\prime }=\epsilon f\left( k_z\right),  \label{modela}
\end{equation}
where $k_z$ and $k_z^{\prime }$ are bounded by the Brillouin zone. Transforming
Eq. (\ref{modela}) to go over to the coordinate representation
we obtain, generally speaking, an integral equation.
The problem consists in the accuracy with which it is possible to obtain a
differential equation in ${\bf r}$ space. Let us consider an equation similar to
Eq. (\ref {modela}),  in which $k_z$ and $k_z^{\prime }$ belong to the entire inverse
space:
\begin{equation}
\int\limits_{-\infty }^{+\infty }H\left( k_z,k_z^{\prime }\right) g\left(
k_z^{\prime }\right)\, {d}k_z^{\prime }=\epsilon g\left( k_z\right) .
\label{modelb}
\end{equation}
The Fourier transform of Eq. (\ref{modelb}) with the system of equations (\ref{all band kp})
taken into account gives a differential equation in the ${\bf r}$ representation.
If the function $g( k_z) $ vanished for $k_z$ not in the Brillouin zone, it would
also be a solution of Eq. (\ref{modela}), and we would solve our problem exactly.
In general this is not so. But in order for Eqs. (\ref{modela}) and (\ref{modelb}) to be
approximately equivalent, it is necessary that $g( k_z)$ be small for $k_z$ not in
the Brillouin zone. In the theory of smooth perturbations this
smallness is ensured by exponentially decaying envelope
functions in the  ${\bf k}$ representation; however, in the case of
discontinuous perturbations the envelope functions are decreasing functions of
$k_z$ with only a power-law falloff. Thus, if the envelope function possesses
one discontinuity, its Fourier transform satisfies
$g( k_z) \propto ( \delta \bar{g}/\bar{g}) \cdot ( k_z) ^{-1}$ for large
$k_z$ (where the exponential contributions associated with the effects of
smooth fields have decayed); where $( \delta \bar{g}/\bar{g}) $ is a typical
relative discontinuity of the function in ${\bf r}$ space. If we consider,
for example, the standard effective-mass equations \cite{kohn-lut} with
discontinuous potentials, then the second derivatives of
the corresponding envelope functions will be discontinuous
with characteristic relative discontinuity of the order of unity
[again, for states whose energies, measured from the band
edge of the left-hand or right-hand material, is of the order of
the band discontinuity], and the error incurred by using differential
equations will be of order $( \bar{k}_z/K ) ^3$, where $K$ is the
radius of the Brillouin zone along the $k _z$ axis.

In the case of a quantum well of width $L$ it is possible to
treat two cases: $\bar{k}_zL$~~{\amsfnt\char"26}~~$1$ and $\bar{k}_zL\ll 1$.
In the first case the error is of the same order as for a single heterojunction; in the
second case it can be of the order $( \bar{k}_zL)^{-1}( \bar{k}_z/K) ^3$. This is
an upper estimate. For a symmetric quantum well in the conduction band, for example,
the error depends on the sign of the product of the values of the envelope function on the
heteroboundaries, and for states of the second 2D subband it is overestimated. In the limiting
case of a narrow quantum well, $L$~~{\amsfnt\char"2E}~~$1/K$, the potential can be replaced
by a $\delta$-function. Then we obtain an envelope function with a discontinuous derivative
and error of order $( \bar{k}_z/K ) ^2$.

Above, in Eq. (\ref{modela}), we tacitly assumed that the Hamiltonian
$H( k_z,k_z^{\prime })$ is defined for all $k_z$ and $k_z^{\prime }$,
belonging to the Brillouin zone. Since our goal is to obtain the one-band
equations, we must take one more circumstance into account. In ${\bf k}$ space near the
${\rm \Gamma }$ point there exists a region $\Lambda _1$ in which
the spectrum of states of the conduction band of the bulk
material can be written in the form of a series in powers of
the quasi-momentum (for states of a degenerate band the
spectrum is determined by a matrix whose elements are such
series). The series converges for $\left| {\bf k} \right| <1/(2 \lambda )$, as follows from
the two-band approximation [again, such an estimate is valid if the effective mass is formed
mainly by the ${\bf k\cdot p}$ interaction]. This is the region described by a Hamiltonian of
the one-band equation. There is also a region $\Lambda _2$, where the interaction of
states of isolated bands with distant bands principally cannot be described by
this series. In our case of sharp heterojunctions the envelope functions in the
${\bf k}$ representation fall off according to a power law; therefore we should also
provide a valid description of the region $\Lambda _2$, which will be done elsewhere in
connection with the problem of inter-valley mixing of states in
heterostructures. Here we only mention that if the ratio of the characteristic band
offset to the energy gap between the states under study in the region $\Lambda _1$ and the
states in the region $\Lambda _2$ is a small parameter $\omega \ll 1$, then the error incurred
by neglecting the region $\Lambda _2$ will be less than or of order $\omega ( a \bar{k}_z ) ^2$.
Thus, the effective radius in ${\bf k}$ space determining the accuracy of reducing the
integral equation to a differential equation is in fact not determined by the size of the
Brillouin zone along the $k _z$ axis, but depends on which bands
are taken to be distant and are ``eliminated'' by the unitary transformation.
In our case, this radius is of the order of $1/{\lambda}$.

Below we will obtain an equation for the conduction
band, leaving out details of the unitary transformation to ${\bf k}$ space and at once
carrying out the transformation to the ${\bf r}$ representation. Formally, the
final equation is a fourth-order differential equation, and the envelope function satisfying it
in the case of a discontinuous potential, the most unfavorable
case for accuracy, has a discontinuous second derivative with
characteristic discontinuity of the order of the second derivative itself.
It is possible to proceed otherwise. Reducing the fourth-order differential equation
to a physically equivalent second-order equation \cite{we1}, we obtain a discontinuous envelope
function with characteristic relative discontinuity of order $( \bar{k}_z \lambda )^2$.
This means that for a single heterojunction or a wide
quantum well the accuracy of the obtained effective-mass
equation is limited as a result of having to take account of all
terms up to order $( \bar{k}_z \lambda ) ^3$ (exclusively). In the case of a narrow quantum
well, on the other hand, even for $L \sim \lambda$ the effective-mass
equation should include only first-order corrections associated with effects
of sharpness of the heterojunction, and considering all remaining orders,
including those leading to spatial dependence of the effective-mass parameters, will yield
an excess of accuracy. In such a case the short-range potential formalism,
already used above to obtain the expansion (\ref{dterms plus}). This will be taken up in
Sec. 4.3.

\section{One-band equations}

\subsection{Conduction band}

\subsubsection{Smooth heterojunction}

The transformation from a many-band system of equations for the envelope functions to a
single-band equation is realized in the standard way \cite{kohn-lut,bir-pikus}. For smooth heterojunctions
the one-band equation for the envelope functions for the $c$ band (conduction band)
was derived in Ref. \cite{we1}. In the ${\bf r}$ representation it has the form:
\[
\epsilon _{c0}{\rm F}_c\left( {\bf r}\right)
+\frac 12m^{\alpha_1} \left( z\right) {\bf p}m^{\beta_1} \left( z\right)
{\bf p}m^{\alpha_1} \left( z\right) {\rm F}_c\left( {\bf r}\right) + 
\Gamma \left( z\right) \Delta U_c {\rm F}_c\left( {\bf r}\right) +
\]
\begin{equation}
+\alpha _0{\bf p}^4{\rm F}_c\left( {\bf r}\right) +\beta _0\left( {\bf p}%
_{||}^2{\rm p}_z^2+{\rm p}_x^2{\rm p}_y^2\right) {\rm F}_c\left( {\bf r}%
\right) + 
\eta \left[ {\bf p\times n}\right] \cdot \sgma \Gamma ^{\prime}\left(
z\right) {\rm F}_c\left( {\bf r}\right) =\epsilon
{\rm F}_c\left( {\bf r}\right).  \label{ema cond a}
\end{equation}
The conduction band offset $\Delta U_c$ and the modified form factor of the
heterojunction $\Gamma ( z) $ are given by
\[
\Gamma \left( z\right) \Delta U_c=G\left( z\right) \delta U_{cc}+{\sum_{%
n}}^{\prime }\frac{\left| \delta U_{cn}\right| ^2}{\epsilon
_{c0}-\epsilon _{n0}}G^2\left( z\right),
\]
so that in all small correctionsÈ $G(z)$ can be replaced by  $\Gamma (z)$.
The position-dependent effective mass is given by
\[
m\left( z\right) =m_1 \left[ 1+m_1\left( \mu _1-\mu _2\right) \Gamma
\left( z\right) \right],
\]
and $m_1$ is the effective mass of the edge of the conduction
band of the left-hand material, and for $m_2$, which is the effective mass
of the conduction band of the right-hand material, we have
$$1/m_2=1/m_1+\mu_2-\mu _1.$$ The parameters $\alpha _1$ and $\beta _1$ are defined
as follows:
\[
\alpha _1 =\frac{\mu _1 }{2\left( \mu _2-\mu _1\right) }, \qquad
2\alpha _1 +\beta _1 =-1.
\]
The parameters $\mu _1$ and $\mu _2$ are given by:
\begin{equation}
\mu _1={\sum_{n}}^{\prime }\frac{2\left|
\left\langle c\right| {\rm p}_x\left| n\right\rangle \right| ^2\delta U_{cc}%
}{m_0^2\left( \epsilon _{c0}-\epsilon _{n0}\right) ^2}-{\sum_{n,l}}
^{\prime }\frac{4\left\langle c\right| {\rm p}_x\left|
n\right\rangle \left\langle n\right| {\rm p}_x\left| l\right\rangle \delta
U_{lc}}{m_0^2\left( \epsilon _{c0}-\epsilon _{n0}\right) \left( \epsilon
_{c0}-\epsilon _{l0}\right) }, \label{mu-1}
\end{equation}
\begin{equation}
\mu _2={\sum_{n,l}}^{\prime }\frac{2\left\langle
c\right| {\rm p}_x\left| n\right\rangle \delta U_{nl}\left\langle l\right| 
{\rm p}_x\left| c\right\rangle }{m_0^2\left( \epsilon _{c0}-\epsilon
_{n0}\right) \left( \epsilon _{c0}-\epsilon _{l0}\right) }. \label{mu-2}
\end{equation}
In Eq. (\ref{ema cond a}) $\alpha _0$ and $\beta _0$ are the nonparabolicity parameters of
the bulk spectrum. Finally, the last parameter entering into the equation is
\[
\eta ={\sum_{n,l}}^{\prime }\frac{{\hbar }^2 \left\langle c\right| 
{\rm p}_z\left| n\right\rangle \left\langle n\right| \left[ \nbla %
\delta U\times {\bf p}\right] _x\left| l\right\rangle \left\langle l\right| 
{\rm p}_y\left| c\right\rangle }{4im_0^4c^2\left( \epsilon _{c0}-\epsilon
_{n0}\right) \left( \epsilon _{c0}-\epsilon _{l0}\right) }.
\]

In the Introduction it was pointed out that an invalid
expression for the effective mass of the edge of the conduction band of a non-basis
semiconductor was obtained in Ref. \cite{burt} by direct elimination of the small
envelope functions. This corresponds to the absence in expression
(\ref{mu-1}) of the second sum, and in expression (\ref{mu-2}) of the terms with $n \ne l$.
Thus it is necessary to be careful when using the method
of direct elimination of small envelope functions.

The relation of the envelope function of the conduction
band ${\rm F}_c( {\bf r}) $ with the total wave function is given by
\[
\Psi \left( {\bf r}\right) =u_{c0}\left\{ 1+2Rm_1\left( \Gamma \left(
z\right) \Delta U_c+\epsilon _{c0}-\epsilon \right) \right\} {\rm F}%
_c\left( {\bf r}\right) +
\]
\begin{equation}
+{\sum_n}^{\prime }\frac{u_{n0}}{\left( \epsilon _{c0}-\epsilon _{n0}\right) 
}\left[ \frac{\hbar \left\langle n\right| {\bf p}\left| c\right\rangle \cdot 
\nbla }{im_0}+\delta U_{nc}\Gamma \left( z\right) +
{\sum_{l,\alpha ,\beta }}^{\prime }\frac{\left\langle n\right| 
{\rm p}_\alpha \left| l\right\rangle \left\langle l\right| {\rm p}%
_\beta \left| c\right\rangle }{m_0^2\left( \epsilon _{c0}-\epsilon
_{l0}\right) }\cdot {\rm p}_\alpha {\rm p}_\beta \right] {\rm F}%
_c\left( {\bf r}\right) .  \label{totbond}
\end{equation}
Here
\[
R=\frac 12{\sum_n}^{\prime }\frac{\left| \left\langle c\right| {\rm p}%
_x\left| n\right\rangle \right| ^2}{m_0^2\left( \epsilon _{c0}-\epsilon
_{n0}\right) ^2},
\]
and the term $2Rm_1( \Gamma ( z) \Delta U_c+\epsilon
_{c0}-\epsilon ) $ inside the braces in
expression (\ref{totbond}) comes from the term $-R{\bf p}^2$ treated as a
perturbation using the standard effective-mass equation. In the
brackets in expression (\ref{totbond}) we neglected the term
\begin{equation}
\frac{\hbar \left\langle n\right| \left[ \nbla U_1 \times {\bf p}\right]
\left| c\right\rangle \cdot \sgma } {4m_0^2c^2}+
\Gamma \left( z\right)\frac{\hbar \left\langle n\right|
\left[ \nbla \delta U \times {\bf p}\right]
\left| c\right\rangle \cdot \sgma } {4m_0^2c^2} , \label{spin-orb}
\end{equation}
since the largest contribution to the matrix elements (\ref{spin-orb}) comes from
the region of the potential near the atomic nuclei
in which the spin-orbit interaction operator can be written
in the form of a product of operators of the electron spin and
the orbital angular momentum, and the function $u_{c0}$ is
spherically symmetric (the orbital momentum is zero).

In Ref. \cite{we1} it was shown that for $d\cdot \bar k_z\ll 1$ Eq. (\ref{ema cond a}) can be
replaced by an equivalent equation where the Heaviside step function $\Theta ( z-z_0) $
replaces the function $\Gamma ( z) $, and the coordinate $z_0$ assigning the position of
the mathematically sharp heterojunction can be chosen arbitrarily within the limits
$-d\leq z_0\leq d$. The method used in Ref. \cite{we1} for this transformation
is not the most convenient. There is a simpler way of
transforming to a mathematically sharp heterojunction based
on the following chain of identities, valid for operators acting on smooth functions:
\[
\Gamma \left( z \right) =\frac 1{2\pi} \int\limits_{-\infty }^{+\infty }\,
{d}k_ze^{ik_z z} \int\limits_{-\infty }^{+\infty }
\Gamma \left( z^{\prime} \right) e^ { -ik_z z^{\prime } }\, {d}z^{\prime} 
=\Theta \left( z-z_0 \right)+
\]
\[
+\frac 1{2\pi} \int\limits_{-\infty }^{+\infty }\,
{d}k_z e^{ik_z \left( z-z_0\right)} \int\limits_{-d }^d \left(
\Gamma\left( z^{\prime}\right) -\Theta\left( z^{\prime}-z_0\right) \right)
e^ { -ik_z \left( z^{\prime }-z_0\right) }\, {d}z^{\prime} \approx
\Theta \left( z-z_0 \right)+
\]
\[
+ \delta \left( z-z_0 \right)\left[ \int\limits_{-d }^d
\Gamma \left( z \right) \, {d}z -\left( d- z_0\right) \right]
+ \delta ^{\prime} \left( z-z_0 \right) \left[ \int\limits_{-d }^d
\Gamma \left( z \right) \left( z_0 -z\right)\, {d}z
+\frac {\left( d-z_0\right)^2}2 \right].
\]
Setting $z_0=0$, instead of Eq. (\ref {ema cond a}) we obtain a more convenient
form of the effective-mass equation for a smooth heterojunction:
\[
\left[ \epsilon _{c0}-\epsilon +\Theta \left( z\right) \Delta U_c
+\Delta U_c \rho _0 \delta \left( z\right) \right]
{\rm F}_c\left( {\bf r}\right) +\frac 12m^{\alpha _2}
\left( z\right) {\bf p}m^{\beta _2}
\left( z\right) {\bf p}m^{\alpha _2} \left( z\right) {\rm F}%
_c\left( {\bf r}\right) + 
\]
\begin{equation}
+\alpha _0{\bf p}^4{\rm F}_c\left( {\bf r}\right) +\beta _0\left( {\bf p}%
_{||}^2{\rm p}_z^2+{\rm p}_x^2{\rm p}_y^2\right) {\rm F}_c\left( {\bf r}%
\right)
+\eta \left[ {\bf p\times n}\right] \cdot \sgma \delta \left( z\right)
{\rm F}_c\left( {\bf r}\right) =0.  \label{ema cond b}
\end{equation}
Here
\[
m\left( z\right) =m_1 \left[ 1+m_1\left( \mu _1-\mu _2\right) \Theta
\left( z\right) \right] , 
\]
and the parameter ${\alpha_2}$ not only depends on the materials of
the heterojunction but also takes into account its finite width
through its dependence on $\Gamma ( z) $:
\[
{\alpha _2} =\frac{\mu _1+2\Delta U_c {\hbar}^{-2}
\left[ d^2 -\int\limits_{-d}^d2\Gamma \left( z\right) z\, {d}z \right] }
{2\left( \mu _2-\mu _1\right) }.
\]
The relation $2{\alpha _2}+{\beta _2} =-1 $ is preserved, and
\[
\rho _0=\int\limits_{-d}^d\Gamma \left( z\right) \, {d}z -d . 
\]
The transformation to a mathematically sharp heterojunction
described above is dictated only by arguments of convenience,
and the explicit form of the function $\Gamma ( z) $ appears
only in integral form in the expressions for the two parameters
$\alpha _2 $ and $\rho _0$.

\subsubsection{Sharp heterojunction}

To include corrections associated with the sharpness of the heterojunction
in Eq. (\ref{ema cond b}), it is necessary in the standard procedure
\cite{kohn-lut,bir-pikus} to take into account the contributions
of the terms $D_{0nn^{\prime }}$, $D_{1nn^{\prime }}$ and ${\bf B}_{0nn^{\prime }}$
in the first-order perturbation theory [${\bf S}_{0cc}=
{\bf S}_{1cc}=0$, as follows from zinc-blende symmetry], and the contributions of the
terms $D_{0nn^{\prime }}$ and ${\bf S}_{0nn^{\prime }}$ should still be treated in the
second-order theory along with the terms $\hbar {\bf k\cdot p}_{nn^{\prime }}/m_0$.
Utilizing symmetry properties, it is not hard to obtain the following additional
term $H_{\rm abr}$ to the Hamiltonian of Eq. (\ref{ema cond a}):
\begin{equation}
H_{\rm abr}=D_{0cc}\delta \left( z\right) +\rho \delta ^{\prime }\left(
z\right) +\tilde{\eta}\left[ {\bf p\times n}\right] \cdot \sgma%
\delta \left( z\right) .  \label{ham abrupt}
\end{equation}
Here
\[
D_{0cc}=-\sum_{j\neq 0}\frac{\left\langle c\mid \delta U\cos \left(
K_jz\right) \mid c\right\rangle }{K_j}\int\limits_{-d}^dG^{%
\prime }(z)\sin \left( K_jz\right) \, {d}z,
\]
\[
\rho ={\sum_{n; j\neq 0}}^{\prime }\frac{\hbar \left\langle c\right| {\rm p}%
_z\left| n\right\rangle \left\langle n\mid \delta U\sin \left(
K_jz\right) \mid c\right\rangle }{i K_jm_0\left( \epsilon
_{c0}-\epsilon _{n0}\right) }\int\limits_{-d}^dG^{\prime }(z)\cos \left(
K_jz\right) \, {d}z+
\]
\[
+\sum_{j\neq 0}\frac{\left\langle c\mid \delta U\cos \left( K_jz\right)
\mid c\right\rangle }{K_j}\int\limits_{-d}^dG^{\prime }(z)\left( 
\frac{\cos \left( K_jz\right) }{K_j}+z\sin \left( K_jz\right)
\right) \, {d}z,
\]
\[
\tilde{\eta}={\sum_{j\neq 0}}\left( {\sum_{n}}^{\prime }\frac{\hbar
\left\langle c\right| \left[ \nbla \left( \sin \left( K_jz\right)
\delta U\right) \times {\bf p}\right] _x\left| n\right\rangle
\left\langle n\right| {\rm p}_y\left| c\right\rangle }
{2 K_jm_0^3c^2 \left( \epsilon _{c0}-\epsilon_{n0}\right) } - \right.
\]
\[
\left. - \frac{\hbar \left\langle c\right| {\nabla}_z \left(
\sin \left( K_jz\right) \delta U\right) \left| c \right\rangle }
{4 K_jm_0^2c^2 } \right)
\int\limits_{-d}^dG^{\prime }(z)\cos \left( K_jz\right)\, {d}z .
\]
Comparing expression (\ref{ham abrupt}) with the Hamiltonian of Eq. (\ref{ema cond b})
we see that for the conduction band taking the sharpness of
the heterojunction into account does not alter the form of the
one-band equation and simply renormalizes the constants
used in it. If we take Eqs. (\ref{ema cond b}) and (\ref{ham abrupt}) into account, the
desired equation for the conduction band takes the form
\[
\left( \epsilon _{c0}-\epsilon +\Theta \left( z\right) \Delta U_c
+d_1 \delta \left( z\right) \right)
{\rm F}_c\left( {\bf r}\right) +\frac 12m^{\alpha} \left( z\right)
{\bf p}m^{\beta}\left( z\right) {\bf p}m^{\alpha } \left( z\right)
{\rm F}_c\left( {\bf r}\right) + 
\]
\begin{equation}
+\alpha _0{\bf p}^4{\rm F}_c\left( {\bf r}\right) +\beta _0\left( {\bf p}%
_{||}^2{\rm p}_z^2+{\rm p}_x^2{\rm p}_y^2\right) {\rm F}_c\left( {\bf r}%
\right)
+d_2 \left[ {\bf p\times n}\right] \cdot \sgma \delta \left( z\right)
{\rm F}_c\left( {\bf r}\right) =0.  \label{ema cond c}
\end{equation}
The term $\rho \delta ^{\prime }( z) $ in expression (\ref{ham abrupt})
was absorbed by the kinetic-energy operator; as a result, for $\alpha$ we obtain
\[
\alpha =\frac{\mu _1+2\Delta U_c {\hbar}^{-2}
\left[ d ^2 -\int\limits_{-d}^d2 \Gamma\left( z\right) z\, {d}z\right]
+4\rho {\hbar}^{-2}}{2\left( \mu _2-\mu _1 \right) }
\]
(here the error incurred in Ref. \cite{we2} has been corrected), and
$2\alpha +\beta =-1 $, with $d_1=D_{0cc} +\Delta U_c \rho _0$ and $d_2 = \eta +\tilde{\eta}$.

Let us discuss the Hamiltonian of Eq. (\ref{ema cond c}). The first term
represents the potential energy of an electron in the $c$ band.
The possible existence of a heterointerface term proportional to $d_1$
was discussed in Ref. \cite{zhu}; it is clear that this
contribution disappears for a mathematically discontinuous
heterojunction (in this rather unrealistic case models (\ref{lattice}) and
(\ref{real lattice}) are identical). The second term is the position-dependent
kinetic-energy operator, which is quadratic in the momentum; such a form was
proposed in Ref.  \cite{morrow} more generally. Note that the parameter
$\alpha$ is not a universal constant but depends both on the materials
and on the shape of the transitional region of the heterojunction, as seems
natural even intuitively (see also Ref. \cite{witold}). If it happens that
$m(z)=const$, i.e., $\mu_1=\mu_2$, indeterminate expressions of the form $1^\infty$ arise
in the form of the kinetic-energy operator used here which
are easily evaluated. Such indeterminate forms do not arise if we use a different,
equivalent form of this operator where we separate out a term analogous to the
relativistic Darwin term (see Ref.~\cite{we1}):
\[
T_2 = {\bf p}\frac 1{2m\left( z\right)} {\bf p} +
\left( \frac {\mu _1\hbar ^2}4 + \Delta U_c \left[ \frac {d ^2}2
-\int\limits_{-d}^d \Gamma\left( z\right) z{\rm d}z\right]
+\rho \right) \delta ^{\prime}\left( z\right).
\]
The third and fourth terms in the Hamiltonian of Eq. (\ref{ema cond c})
describe corrections to the weak nonparabolicity and depend
only on the bulk parameters. The fifth term describes the
interface spin-orbit interaction (see, e.g., Ref. \cite{vasko}), whose strength
($d_2$) depends not only on the materials of the heterojunction but also on the
shape of the transitional region. The relation between the total wave function and the
envelope functions of the conduction band is given by Eq. (\ref{totbond}), where we can
set $\Gamma (z)=\Theta (z)$.

\subsubsection{Boundary conditions on the envelope function}

From Eq. (\ref{ema cond c}) it is not hard to obtain boundary conditions
imposed on the envelope function at the heterojunction; to this end it is necessary to
reduce Eq. (\ref{ema cond c}) to a second-order differential equation by employing the
smallness of the contribution proportional to ${\bf p}^4$ (see Ref. \cite {we1}). We
present only the result:
\[
\left( \begin{array}{c}
{\rm F}_c\left( {\bf r}\right)\\
{\rm F}^{\prime}_c\left( {\bf r}\right)
\end{array} \right) \Biggr| _{z=+0}=
\left(
\begin{array}{cc}
d_{11} & 0 \\ d_{21} & d_{22} \end{array} \right)
\left( \begin{array}{c}
{\rm F}_c\left( {\bf r}\right)\\
{\rm F}^{\prime}_c\left( {\bf r}\right)
\end{array} \right) \Biggr| _{z=-0} ,
\]
where
\[
d_{11} = 1 +4 m_1^2\alpha_0 \Delta U_c +
m_1 \alpha \left( \frac 1{m_2} -\frac 1{m_1} \right),
\]
\[
d_{21} = \frac{2m_1}{\hbar^2} \left( d_1 +d_2
\left[ {\bf p\times n}\right] \cdot \sgma \right),
\]
and
\[
d_{22} = \frac{m_2}{m_1} +4 m_1^2\alpha_0 \Delta U_c -
m_1 \alpha \left( \frac 1 {m_2} -\frac 1{m_1} \right).
\]

In the approximation of the standard effective-mass
method we have a position-independent effective mass and a
discontinuous (step-function) potential. Corrections which
are first order in the small parameter of the problem are taken
into account by including in the standard equation a $\delta$-function potential
proportional to $d_1$ (formally, this is a correction of order $d \cdot \bar{k}_z$).
The complete equation (\ref{ema cond c}) also includes all corrections of order
$(\lambda \bar{k}_z)^2$. It is not possible to take into account smaller contributions
in the one-band version of the method of envelope functions because the
unavoidable error arising in the transformation from the many-band system of integral
equations to the one-band differential equation is of the same order.

It is not hard to generalize the above results to the case
of an arbitrary heterostructure. It is easy to do this proceeding from symmetry
arguments for a symmetric quantum well with two equivalent heterojunctions with coordinates
$z=0$ and $z=L$, where $L={\cal N}a/2$, and $\cal N$ is an integer. The effective Hamiltonian is
\[
H^{el}=\epsilon _{c0} + \Delta U_c \{ \Theta \left( z\right)
-\Theta \left( z-L \right) \}
+\frac 12m^{\alpha}\left( z\right) {\bf p}m^{\beta}
\left( z\right) {\bf p}m^{\alpha}\left( z\right) +\alpha _0{\bf p}^4+ 
\]
\[
+\beta _0\left( {\bf p}_{||}^2{\rm p}_z^2+{\rm p}_x^2{\rm p}_y^2\right)
+d_1\{ \delta \left( z\right)+\delta \left( z-L\right) \}
+d_2\{ \delta \left( z\right) -\delta \left( z-L \right) \}
\left[ {\bf p\times n}\right] \cdot \sgma .
\]

\subsection{Valence band}

The distinction between the effective-mass method for
the valence band and the effective-mass method for the conduction band consists,
in principle, simply of the necessity of considering more elaborate equations
in the case of the valence band. The main points of the problem of deriving the
equation for the hole states with position-dependent effective-mass parameters were
noted in Ref. \cite{we2}. The equation for the $c$ band already contains three new parameters
which depend on the bulk properties of the materials of the heterostructure and on the
properties of the heteroboundary. For the valence band, there is a larger number of such
parameters, which may be seen as rendering such an equation of little practical value.
Therefore we limit ourselves here to a derivation of first-order corrections to the standard
equation in the above small parameter.

In the basis  $\{ \left| J, j_z\right\rangle \}$ of eigenfunctions of the total angular
momentum $J$ and its projection $j_z$, with these eigenfunctions being combinations of
the Bloch functions $\cal X$, $\cal Y$ and $\cal Z$ of the top of the valence band
(transforming according to the representation $\rm \Gamma_{15}$) and the spin,
the matrix of the effective Hamiltonian for the valence band ${\bf H}$ in this approximation
is a sum of the $6\times 6$ matrix of the standard kinetic energy operator ${\bf T}$
(we neglect the small contribution of ${\bf k}$-linear bulk terms from the spin-orbit
interaction) and the $6\times 6$ matrix of the potential energy operator ${\bf V}$:
\[
{\bf H} = {\bf T} + {\bf V}.
\]
Of course, in the standard approximation ${\bf V}$ contains only
diagonal discontinuous (step-function) potentials. Additional terms appear in the
approximation that follows which are diagonal and non-diagonal $\delta$-function potentials.

It is convenient first to find the elements of the potential energy matrix
$\tilde {\bf V}$ in the basis \{$\cal X$, $\cal Y$, $\cal Z$\}, and then compose
from them the necessary linear combinations and transform to ${\bf V}$.
\[
\tilde {\rm V}_{\cal ZZ}=\tilde {\rm V}_{\cal XX}=\epsilon _{{\cal X}0}
+\delta U_{\cal XX} G\left( z\right) +D_{0\cal XX} \delta \left( z\right)
\approx
\]
\[
\approx \epsilon _{{\cal X}0} +\delta U_{\cal XX} \Theta \left( z\right)
+\left( D_{0\cal XX} +\rho _0\delta U_{\cal XX}\right)\delta\left( z\right);
\]
\[
\tilde {\rm V}_{\cal XY}
=\frac 1{3i}\left( \Delta +\delta\Delta G\left( z\right) \right)\sigma _z
+D_{0\cal XY} \delta \left( z\right)
+S_{0\cal XY}^z \delta \left( z\right) \sigma_z \approx
\]
\[
\approx\frac 1{3i}\left( \Delta +\delta\Delta\Theta\left( z\right) \right)
\sigma _z +D_{0\cal XY}\delta \left( z\right)
+\left( S_{0\cal XY}^z +\frac {\rho _0\delta\Delta}{3i} \right)
\delta \left( z\right)\sigma_z ,
\]
where
\[
\Delta =\frac{3\hbar i\left\langle {\cal X}\right| \left[ \nbla %
U_1\times {\bf p}\right]_z \left| {\cal Y}\right\rangle }
{4m_0^2c^2}, \quad
\delta\Delta =\frac{3\hbar i\left\langle {\cal X}\right| \left[ \nbla %
\delta U\times {\bf p}\right]_z \left| {\cal Y}\right\rangle }
{4m_0^2c^2}.
\]
Analogously,
\[
\tilde {\rm V}_{\cal XZ}
=\frac i3\left( \Delta +\delta\Delta G\left( z\right) \right)\sigma _y
+S_{0\cal XZ}^y \delta \left( z\right) \sigma_y \approx
\]
\[
\approx\frac i3\left( \Delta +\delta\Delta\Theta\left( z\right) \right)
\sigma _y +\left( S_{0\cal XZ}^y + \frac {i\rho _0\delta\Delta}{3}
\right) \delta \left( z\right) \sigma _y .
\]
The remaining elements of $\tilde {\bf V}$ can be obtained from those
shown above by cyclic permutation of the indices.

The contribution of the sixth and seventh terms on the left-hand side of
Eq. (\ref{all band kp}), as can be seen, is not included in $\tilde {\bf V}$,
since it is negligibly small for the following reason:
\[
\left\langle {\cal X}\mid {\nabla }_z U_1 \mid {\cal Y}\right\rangle =0
\]
(see Ref. \cite {bir-pikus}, Sec. 21). This means that the matrix element
\[
\left\langle {\cal X}\mid {\nabla }_z \delta U \mid {\cal Y}\right\rangle =
\left\langle {\cal X}\mid {\nabla }_z U_2 \mid {\cal Y}\right\rangle .
\]
For the Bloch functions of the $n$th band at the $\rm \Gamma$ point of the
right-hand crystal in the non-relativistic limit ($\tilde u_{n0}$) we have:
\begin{equation}
\tilde u_{n0}=u_{n0}+{\sum_{n^{\prime }}}^{\prime }
\frac{\delta U_{n^{\prime }c}u_{n^{\prime }0}}
{\epsilon _{c0}-\epsilon _{n^{\prime }0}}, \label{left-right}
\end{equation}
and for the corresponding functions of the edge of the valence band of the right-hand
crystal $\tilde {\cal X}$ and $\tilde {\cal Y}$ the relation $\left\langle
\tilde {\cal X}\mid {\nabla }_z U_2 \mid \tilde {\cal Y}\right\rangle=0$ holds.
That is, the seventh term on the left-hand side of Eq. (\ref{all band kp}) gives
corrections only of order $(\lambda \bar{k}_z)^3$, not $\lambda \bar{k}_z$. In our
approximation it is necessary in general to neglect the difference between the Bloch functions
$\tilde u_{n0}$ and $u_{n0}$. Hence it follows from invariance of the equation under
time reversal that the contribution from the sixth term on the left-hand side of
Eq. (\ref{all band kp}) is negligibly small.

It is not difficult now to obtain the elements of ${\bf V}$. We
choose the phases the same as was done in Ref. \cite{kohn-lut} and write
\[
\left| 1\right\rangle \equiv \left| \frac 32,\frac 32\right\rangle , \quad
\left| 2\right\rangle \equiv \left| \frac 32,-\frac 32\right\rangle , \quad
\left| 3\right\rangle \equiv \left| \frac 32,\frac 12\right\rangle ,
\]
\[
\left| 4\right\rangle \equiv \left| \frac 32,-\frac 12\right\rangle , \quad
\left| 5\right\rangle \equiv \left| \frac 12,\frac 12\right\rangle , \quad
\left| 6\right\rangle \equiv \left| \frac 12,-\frac 12\right\rangle
\]
(pairs of states of heavy holes, light holes, and states of the
split-off band). Thus, the desired potential-energy matrix
takes the form
\begin{eqnarray}
{\bf V}&=&\left( \matrix{
V_{\Gamma_8} {\bf 1}&V_0 \sigma_y&-i\sqrt 2 V_0\sigma_y\cr
V^{\dagger}_0\sigma_y&V_{\Gamma_8}{\bf 1}&{\bf 0}\cr
i\sqrt 2 V^{\dagger}_0\sigma_y&{\bf 0}&V_{\Gamma_7}{\bf 1} } \right),
\label{valpotmatrix}
\end{eqnarray}
where
\[
{\rm V}_{\Gamma_8}=E_{\Gamma_8}+\Delta U_{\Gamma_8}\Theta
\left( z\right)+\chi_1 \delta \left( z\right) ,
\]
\[
{\rm V}_{\Gamma_7}=E_{\Gamma_7}
+\Delta U_{\Gamma_7}\Theta \left( z\right)+\chi_2 \delta \left( z\right) ,
\]
\[
{\rm V}_{0}={D_{0{\cal XY}}\over \sqrt{3}}\delta \left( z\right) .
\]
Here we have introduced the notation
\[
E_{\Gamma_8}=\epsilon_{{\cal X}0}+\frac 13\Delta , \quad
\Delta U_{\Gamma_8}=\delta U_{\cal XX} +\frac 13\delta \Delta ,
\]
\[
E_{\Gamma_7}=\epsilon_{{\cal X}0}-\frac 23\Delta , \quad
\Delta U_{\Gamma_7}=\delta U_{\cal XX} -\frac 23\delta \Delta ,
\]
\[
\chi_1 =D_{0\cal XX} +\rho _0\delta U_{\cal XX}
+i S_{0\cal XY}^z +\frac 13\rho _0\delta\Delta ,
\]
\[
\chi_2 =D_{0\cal XX} +\rho _0\delta U_{\cal XX}
-2i S_{0\cal XY}^z -\frac 23\rho _0\delta\Delta .
\]
The expressions for $D_{0\cal XX}$, $D_{0\cal XY}$ and $S_{0\cal XY}^z$ have the
following form:
\[
D_{0\cal XX}=-\sum_{j\neq 0}\frac{\left\langle {\cal X}\mid \delta U
\cos \left( K_jz\right) \mid {\cal X}\right\rangle }{K_j}
\int\limits_{-d}^dG^{\prime }(z)\sin \left( K_jz\right) \, {d}z,
\]
\begin{equation}
D_{0\cal XY}=\sum_{j\neq 0}
\frac{\left\langle {\cal X}\mid \delta U\sin \left( K_jz\right)
\mid {\cal Y}\right\rangle }{K_j}\int\limits_{-d}^dG^{\prime }(z)\cos
\left( K_jz\right) \, {d}z , \label{d0xy}
\end{equation}
\[
S_{0\cal XY}^z= -\sum_{j\neq 0}\frac{\hbar \left\langle {\cal X}\right|
\left[ \nbla \left( \cos \left( K_jz \right) \delta U\right) \times {\bf p}
\right]_z \left| {\cal Y}\right\rangle }{4 K_jm_0^2c^2}
\int\limits_{-d}^dG^{\prime }(z) \sin \left( K_jz \right)\, {d}z.
\]

Thus, within the framework of the ${\bf k\cdot p}$ method we have
shown that in (001) III-V heterostructures mixing of heavy
($hh$) and light ($lh$) holes takes place at the center of the 2D
Brillouin zone (see Ref. \cite{ivchenko} and the references cited therein) 
which bears no relation to the ${\rm k}_z$-linear bulk terms from the
spin-orbit interaction. This mixing is governed by the parameter $D_{0\cal XY}$,
which was estimated in Ref. \cite{ivchenko} on the basis
of experimental data for GaAs/AlAs heterostructures: $D_{0\cal XY}\simeq 500$ {\rm meV \AA }.

In Ref. \cite{we2} it was concluded that the strength of the mixing of the heavy
and light holes at the center of the 2D Brillouin zone is greater for sharp
heterojunctions than for heterojunctions with smoothly varying chemical
composition. But this is valid only in model (\ref{lattice}).  Generally speaking,
one can draw conclusions only about the dependence of the strength of this mixing on
the structure of the transitional region of the heterojunctions.

For a symmetric quantum well with boundaries $z=0$ and $z=L$ the elements of the
potential energy matrix can be easily obtained from symmetry arguments:
\[
{\rm V}_{\Gamma_8}=E_{\Gamma_8} +\Delta U_{\Gamma_8}
\left(\Theta \left( z\right)-\Theta \left( z-L\right) \right)
+\chi_1 \left(\delta \left( z\right)+\delta \left( z-L\right) \right),
\]
\[
{\rm V}_{\Gamma_7}=E_{\Gamma_7}+\Delta U_{\Gamma_7}
\left( \Theta \left( z\right)-\Theta \left( z-L\right) \right)
+\chi_2 \left( \delta \left( z\right) +\delta \left( z-L\right) \right) ,
\]
\[
{\rm V}_{0}={D_{0{\cal XY}}\over \sqrt{3}}
\left( \delta \left( z\right)-\delta \left( z-L\right) \right) .
\]

\subsection{Equation for the envelope function for a narrow quantum well in
the conduction band}

We will devote separate attention to the problem of electron states in narrow
quantum wells because at present it is widely held that the effective-mass method is
inapplicable in such case. Here we treat only states of the $c$ band in a III-V (001)
heterostructure consisting of related semiconductors of isolated-quantum-well-type (or
narrow-barrier-type) for the case when its width satisfies $L$~{\amsfnt\char"2E}~$\lambda$.
Hole states are easily treated in an analogous way. Now the contributions from the sharpness
of the potential begin to play a much greater role than in the case of a wide quantum well.
In fact, the estimate  $\bar{k}^2_z/2m \sim \delta U_{cc}$ is valid only when the width
of the quantum well is greater than or of the order of the characteristic localization
length of the states. For states in a narrow quantum well, on the other hand, of course
$\bar{k}^2_z/2m \sim \delta U_{cc} \bar{k}_zL$, and the contribution to the energy eigenvalue
from the terms associated with sharpness of the potential can be estimated to first order
as $\delta U_{cc}\bar{k}_za$, which implies that they must be taken into account even in the
zeroth approximation.

Thus, the potential of the heterostructure under consideration can be written as
\[
U=U_1+P\left[ U_2-U_1\right] \equiv U_1+P\left( z\right) \delta U,
\]
where $U_1$ and $U_2$ are the periodic crystal potentials of the ``barrier'' (basis) and
``well'' semiconductor, respectively, and $P( z )$ is the form factor of the heterostructure.
We choose $L$ so that $P( z<-L/2) = P( z>L/2) =0$. It is natural to consider
$P( z)$ as a local function on the scale of variation of the envelope function of the
conduction band since $L \bar{k}_z \ll 1$. The one-band equation has the following form:
\begin{equation}
\left[ \epsilon _{c0}-\epsilon +\frac{{\bf p}^2}{2m }
+b_1\delta \left( z\right) +b_2{\delta}^{\prime}\left( z\right)
+b_3\delta \left( z\right)\left[ {\bf p\times n}\right] \cdot \sgma
\right]{\rm F}_c\left( {\bf r}\right) =0, \label{p21}
\end{equation}
where $m$ is the position-independent effective mass, and we
have the following expressions for the three parameters $b_i$ ($i=1,\ 2,\ 3$):
\begin{equation}
b_1=\sum_{j}\left\langle c\mid \delta U\cos \left(
K_jz\right) \mid c\right\rangle \int\limits_{-L/2}^{L/2}
P(z)\cos \left( K_jz\right) \, {d}z; \label{b1}
\end{equation}
\[
b_2={\sum_{n; j}}^{\prime }\frac{\hbar \left\langle c\right| {\rm p}%
_z\left| n\right\rangle \left\langle n\mid \delta U\sin \left(
K_jz\right) \mid c\right\rangle }{i m_0\left( \epsilon
_{c0}-\epsilon _{n0}\right) }\int\limits_{-L/2}^{L/2}P(z)\sin \left(
K_jz\right) \, {d}z-
\]
\[
-\sum_{j}\left\langle c\mid \delta U\cos \left( K_jz\right)
\mid c\right\rangle \int\limits_{-L/2}^{L/2}P(z)\cos
\left( K_jz\right) z\, {d}z;
\]
\[
b_3={\sum_{j}}\left( {\sum_{n}}^{\prime }\frac{\hbar
\left\langle c\right| \left[ \nbla \left( \sin \left( K_jz\right)
\delta U\right) \times {\bf p}\right] _x\left| n\right\rangle
\left\langle n\right| {\rm p}_y\left| c\right\rangle }
{2 m_0^3c^2 \left( \epsilon _{c0}-\epsilon_{n0}\right) } - \right.
\]
\[
\left. - \frac{\hbar \left\langle c\right| {\nabla}_z \left(
\sin \left( K_jz\right) \delta U\right) \left| c \right\rangle }
{4 m_0^2c^2 } \right)
\int\limits_{-L/2}^{L/2}P(z)\sin \left( K_jz\right)\, {d}z ,
\]
in which the summation index $j$ includes zero (the terms with $j=0$
represent the contribution of the smooth part of the potential). The term in the potential
energy proportional to $b_1$ gives the main contribution, and the two remaining terms
are corrections of order $\lambda \bar{k}_z $. For a symmetric structure, $P(z)=P(-z)$,
the equation simplifies: $b_2=b_3=0$.

Strictly speaking, Eq. (\ref{p21}) is invalid from a mathematical point of view
(it does not have a rigorous nontrivial solution); therefore, it needs to be put
into a different form using the smallness of the term proportional to $b_2$, and discarding
corrections of order $(\lambda \bar{k}_z )^2$. It is necessary to invoke the
approximate relation
\begin{equation}
b_1\delta \left( z\right) +b_2{\delta}^{\prime}\left( z\right) \approx
b_1\delta \left( z+\frac{b_2}{b_1}\right), \label{p22}
\end{equation}
and in the term proportional to $b_3$, $\delta (z)$ can be replaced by $\delta (z+b_2/b_1)$
for simplicity. We obtain the valid equation
\begin{equation}
\left[ \epsilon _{c0}-\epsilon +\frac{{\bf p}^2}{2m }
+b_1\delta \left( z+\frac{b_2}{b_1}\right)
+b_3\delta \left( z+\frac{b_2}{b_1}\right)
\left[ {\bf p\times n}\right] \cdot \sgma \right]
{\rm F}_c\left( {\bf r}\right) =0. \label{p23}
\end{equation}
Note that for states of one band (if we are not interested, for
example, in inter-band transitions) in a heterostructure with one narrow layer the
value of $b_2$ does not play a role: in Eq. (\ref{p21}) we can shift the origin
$z^{\prime}=z+b_2/b_1$.

Assume now that we are dealing with a structure containing two narrow layers lying near
one another, such that the distance between them is of the order of $\lambda$.
In this case, an upper estimate on the error arising from the transformation to
a one-band differential equation is $\lambda \bar{k}_z $
(this is valid, in particular, in the case when the constant $b_1$ describing the
potential of the first layer is equal to the constant describing the
potential of the second layer taken with opposite sign). Then
we should remove from consideration those terms containing $b_2$ and $b_3$ in Eq. (\ref{p21}).

One could probably treat both layers as one local perturbation, thereby decreasing the error,
and obtain Eq. (\ref{p21}) with one set of parameters of the local perturbation.
But in the situation for which we obtained an upper estimate of the error, however,
one could not then guarantee the smallness of the parameter
$\bar{k}_z \tilde {b_2}/\tilde {b_1}$ on which Eq. (\ref{p22}) is based. If it is not small,
then one could not say that Eq. (\ref{p22}) is mathematically correct, which would
imply the inapplicability of such an approach.

Thus, taking the above modification into account, we can
also apply the effective-mass method to electron states in
heterostructures with super-thin layers. In this regard the situation
can arise in which the potential of a thick layer of some
semiconductor plays the role of a barrier for the electron
states while a thin layer of the same material will couple
states and vice versa, depending on the sign of the parameter $b_1$. The sign of this parameter,
as can be seen from Eq. (\ref{b1}), can be different from the sign of the parameter
$\left\langle c\mid \delta U\mid c\right\rangle$, which defines the conduction
band offset at the heterojunction.

\section{Hierarchy of effective-mass equations and discussion of results}

We have derived a many-band ${\bf k\cdot p}$ system of integral
equations (\ref{all band kp}) which can be used to describe electron states
in heterostructures with atomically sharp variation of their
chemical composition. The system contains contributions in
the form of converging power series in $(k_z-k_z^{\prime })$ which are due
to the sharpness of the heterojunction. For example, such
terms were discarded in Refs. \cite{burt}, \cite{foreman1}, and \cite{foreman2}
so that the effects of a discontinuous change in the crystal potential of
the structure near a heterojunction were in fact neglected. It
is specifically the presence of these terms that distinguishes
the many-band system of ${\bf k\cdot p}$ equations derived here from
the system obtained by Leibler \cite{leibler} for heterostructures with
smooth heterojunctions. All of the papers known to us which
use one-band or many-band schemes of the method of envelope functions to
describe electron states in heterostructures apply the ${\bf k\cdot p}$ system of Leibler.
Very often the distinction between the Bloch functions for the component semiconductors
of the heterostructure is also neglected, which gives, in particular,
 $\delta U_{nn^{\prime }}=0$ for $n \neq n^{\prime }$.
Taking into account the terms due to sharpness of the heterojunction, the
${\bf k\cdot p}$ method can also be used to describe inter-valley mixing of states in
heterostructures, including the problem of $\rm \Gamma $-$\rm X_z$ mixing of
states in (001) heterostructures.

The main limitation on the accuracy of the method of envelope functions
employing differential equations is the procedure of transforming from
${\bf k}$ to ${\bf r}$ space. The one-band differential equation (of fourth order) with
position-dependent effective mass is valid for structures with characteristic width of
layers much greater than $\lambda $, where the length $\lambda $ was defined in Sec. 2.

Above we considered in detail how taking account of contributions of the sharpness of
the interface potential modifies the equation for states of the conduction band in
(001) heterostructures of related, lattice-matched III-V semiconductors, derived in
Ref. \cite{we1}. Formally, the resulting equation for a sharp heterojunction differs
from that for a smooth heterojunction only by renormalization of the parameters entering
into it. In the case of the valence band, on the other hand, taking account of the
sharpness of the heterojunction leads to qualitatively new effects (mixing of heavy and light
holes for ${\bf k}_{||}=0$).

For heterostructures with wide layers it is possible to construct a hierarchy of
approximations of the one-band method of envelope functions according to the parameter
$\lambda \bar{k}_z$, where $\bar{k}_z$ is the characteristic value of the
quasi-momentum of the state. For example, for an isolated heterojunction
we have the following.

0) Zeroth level of the hierarchy for electrons. In the effective-mass approximation in
which small corrections in the order parameter have been neglected, we have
the usual equation with position-independent effective mass and a discontinuous (step-function)
potential:
\begin{equation}
\left( \epsilon _{c0}+\Theta \left( z\right) \Delta U_c
+\frac {{\bf p}^2}{2m} \right) {\rm F}_c\left( {\bf r}\right)
=\epsilon {\rm F}_c\left( {\bf r}\right). \label{zerolevel}
\end{equation}

1) First level of the hierarchy. First-order corrections are taken into account (here, in fact,
the small parameter $d \cdot \bar{k}_z$ plays a role; $2d$ is
the width of the transitional region of the heterojunction) by including a
$\delta$-function potential in Eq. (\ref{zerolevel}), which is localized
at the heteroboundary:
\[
\left( \epsilon _{c0}+\Theta \left( z\right) \Delta U_c
+d_1\delta \left( z\right) +\frac {{\bf p}^2}{2m} \right)
{\rm F}_c\left( {\bf r}\right)=\epsilon {\rm F}_c\left( {\bf r}\right),
\]
where $d_1$ is given by the complicated expression in Sec. 4.1.

2) Second level of the hierarchy. Equation (\ref{ema cond c}) includes all corrections of
order $(\lambda \bar{k}_z)^2$. Smaller contributions, of third order and higher, cannot be
taken into account correctly in a one-band version of the method.

For hole states we obtained, see Eq. (\ref{valpotmatrix}), the first-order
corrections to the standard equation [that is, the first
step of the hierarchy of effective-mass equations for the holes] and showed that for (001)
heterostructures mixing of heavy and light holes at the center of the 2D Brillouin zone
does indeed take place and that contributions from the sharpness of the heterojunction
potential determine the strength of this mixing so that it depends on the microscopic
structure of the heteroboundaries. In Ref. \cite{foreman2} it is asserted, in particular,
that such mixing of heavy and light holes is caused by a difference in the Bloch
functions for the component semiconductors of the heterostructure and it is absent,
if one neglects such a difference, or, what should be equivalent, if the
Bloch functions of all the bulk semiconductors comprising the structure are the same
set of functions. If this is indeed the case, then the contribution of these terms [even without
taking symmetry arguments into account] would be only of order
$(\lambda \bar{k}_z)^2$, as can be seen from Eq. (\ref{left-right}). In fact, however,
the difference in the Bloch functions does not play a substantial role.
To prove this, consider the idealized situation of a (001) homojunction---the problem of
hole states in a weak but not smooth external potential, say, $W ( z ) = G( z ) W_0$,
where $W_0$ is a constant assigning the jump of the potential, small in comparison with
the band gap. In this case the point symmetry of the structure, $\rm C_{2v}$,
also admits the existence of mixing of heavy and light holes at the center of the 2D
Brillouin zone \cite{ivchenko}, and for the coefficient governing this mixing, $D _{0\cal XY}$,
instead of formula (\ref{d0xy}) we have
\[
D_{0\cal XY} =\sum_{j\neq 0}
\frac{W _0\left\langle X\mid \sin \left( K_jz\right)
\mid Y\right\rangle }{K_j}\int\limits_{-d}^dG^{\prime }(z)\cos
\left( K_jz\right) \, {d}z.
\]
This is direct proof of our assertion.

The independent parameters $\alpha$, $d_1$ and $d_2$ introduced in
the present work and appearing in Eq. (\ref{ema cond c}), and also $\chi_1$, $\chi_2$ and
$D_{0\cal XY}$ entering into the effective potential energy operator for the
valence-band states, depend not only on the bulk properties of the materials of the
heterojunction, but also on its microscopic structure. All these parameters determine the
heterointerface contribution to the potential energy. At the same time, it is well known
\cite{dipole}, that as a consequence of the
possible appearance of an electric dipole moment at the interface of two materials the
magnitude of the potential jump at the interface can also depend on the microscopic structure
of the boundary. This is not described in our model of a heterojunction because we do
not take into account the effect of such a dipole. Including the corresponding discontinuous
electrostatic potential in Eq. (\ref{lattice}) and developing it according
to the scheme laid out above also yields the desired effect.

The electron states in heterostructures consisting of thin
layers whose thickness is less than or of the order of $\lambda$ can
be treated only in the approximation quadratic in the momentum operator.
In this case an account of terms due to the sharpness of the potential becomes
necessary already in the zeroth approximation. This is clear from Eqs. (\ref{b1}).
In this regard, the following situation is possible: as the width of the
quantum well is decreased, the bound state can disappear or
conversely, a thin layer---nominally a ``barrier'' layer---of some semiconductor can
create an attractive potential and form a bound state.
It is possible that just such a situation was observed in Ref. \cite{schwabe}
and then modeled in Ref. \cite{ventra}.

\section*{Acknowledgements}

\noindent
The authors are grateful to E.~L.~Ivchenko and A.~A.~Gorbatsevich for
helpful discussions of a number of results of this work.
This work was carried out with the financial
support of the Russian Fund for Fundamental Research
(Grant No. 99-02-17592) and INTAS-RFBR (Grant No. 97-11475), and also the
Russian Ministry of Science under the auspices of the programs
``Physics of Solid-State Nanostructures'' (Grant No. 99-1124) and
``Atomic Surface Structures'' (Grant No. 3.1.99).

\section*{Appendix A: The Dirac equation with position-va\-ri\-ant gap}

Let us consider a model Dirac equation with a position-dependent gap
$2m ( {\bf r} ) c^2$
\[
\left[
\begin{array}{cc}
m \left( {\bf r} \right) c^2 & c {\bf \sgma \cdot p} \\
c{\bf \sgma \cdot p} & -m \left( {\bf r} \right) c^2
\end{array}
\right] \left( 
\begin{array}{c}
\varphi _e \\
\varphi _p 
\end{array}
\right) =\epsilon \left( 
\begin{array}{c}
\varphi _e \\ 
\varphi _p 
\end{array}
\right) , 
\]
where $\varphi _e$ and $\varphi _p$ are the electron and positron components
of the wave function, respectively. Let $m ( {\bf r} )$ vary weakly in
space, i.e., $m ( {\bf r} )=\tilde m + \delta m ( {\bf r} )$, so that
$\delta m ( {\bf r} ) / \tilde m \ll 1$.
With the help of a Foldy-Wouthuysen unitary transformation it is
quite simple, for example, following the scheme laid out in Ch. 20, Sec. 33
of Ref. \cite{messia}, to obtain a one-band equation describing the states of the
electron. Thus, the equation in which all small terms have been discarded is the ordinary
Schr\"odinger equation
\[
\left( m \left( {\bf r} \right) c^2 + \frac{{\bf p}^2}{2 \tilde m} \right)
\tilde \varphi _e = \epsilon \tilde \varphi _e,
\]
where $\tilde \varphi _e$ is the transformed electron wave function. The
equation, on the other hand, in which all terms of higher order than
$\delta m ( {\bf r} ) / \tilde m$ have been neglected has the form
\[
H \tilde \varphi _e = \epsilon \tilde \varphi _e,
\]
$$
H = m \left( {\bf r} \right) c^2 + \frac{{\bf p}^2}{2 \tilde m} -
\frac{{\bf p} \delta m \left( {\bf r} \right) {\bf p}} {2 {\tilde m}^2}+
\frac{ {\hbar}^2 {\nbla}^2 \delta m \left( {\bf r} \right) } {8 {\tilde m}^2}-
\frac {{\bf p}^4} {8 {\tilde m}^3 c^2}-
\frac {\hbar \left[ \nbla \delta m \left( {\bf r} \right) \times 
{\bf p}\right] \cdot \sgma } {4 {\tilde m}^2}. \eqno\hbox{(A1)}
$$
\begin{figure} [ht]
\epsfig{figure=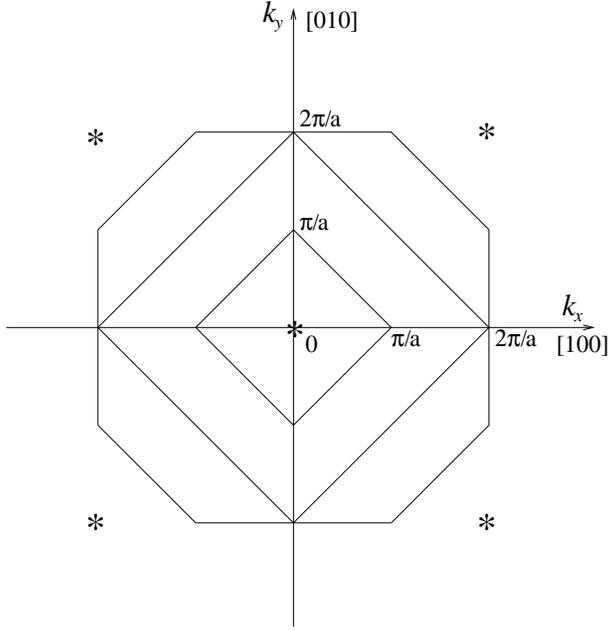,width=8.cm}
\caption{Projection of the bulk Brillouin zone (the region bounded by the
octagon) and sites of the inverse lattice (denoted by asterisks) onto the (001) plane.
The square with diagonal $2\pi/a$ bounds the region where there are no
2D transfer processes.} \label{BZ}
\end{figure}
All terms in Eq. (A1) with the exception of the third which
describes the position-dependent mass can be taken as ``ordinary.''
The second, third, and fourth terms (the fourth is the Darwin term) can either be written
in the following form:
\[
T_{2} = \frac{1} {2} {\bf p} \frac{1} {m \left( {\bf r} \right)} {\bf p}+
\frac{{\hbar}^2 {\nbla}^2 \delta m \left( {\bf r} \right) } {8 {\tilde m}^2},
\]
or combined into one, quadratic in the momentum, kinetic energy operator:
\[
T_2 = \frac{1} {2} \frac{1} {\root 4 \of {m \left( {\bf r} \right)}}
{\bf p} \frac{1} {\sqrt{m \left( {\bf r} \right)}} {\bf p}
\frac{1} {\root 4 \of {m \left( {\bf r} \right)}},
\]
or some other equivalent form can be used. For example,
Ref. \cite{cavalkante} uses the following form for $T_2$:
\[
T_2 = \frac{1} {4} \left[ {\bf p} \frac{1} {\sqrt{m \left(
{\bf r} \right)}} {\bf p} \frac{1} {\sqrt{m \left( {\bf r} \right)}} +
\frac{1} {\sqrt{m \left( {\bf r} \right)}} {\bf p}
\frac{1} {\sqrt{m \left( {\bf r} \right)}} {\bf p} \right] .
\]
Thus, in the model Dirac equation with position-dependent
gap the concept of a position-dependent effective mass
shows up only beyond the frames of the non-relativistic (qua\-dra\-tic) approximation.

\section*{Appendix B: Regarding transfer processes in the two-dimensional Brillouin zone for a
(001) he\-te\-ro\-struc\-tu\-re}

Let us consider the second sum in Eq. (\ref{sums}) describing
transfer processes in the two-dimensional Brillouin zone and
prove that it does not contribute in the case of interest to us
of states near the Brillouin zone center in (001) heterostructures.
Since the function ${\cal G}(q)$ is nonzero for any $q$, there also
exist nonzero vectors of the inverse lattice ${\bf K}_j$ for which
${\bf k}_{||}^{\prime}-{\bf k}_{||}={\bf K}_{||j}$. Here ${\bf k}_{||}^{\prime}$ and
${\bf k}_{||}$ are components of vectors belonging to the Brillouin zone.
There exits a finite number of vectors ${\bf K}_j$ possessing this property.
Therefore, in general we should also retain the second sum in expression (\ref{sums}).
Let us now consider the interesting case of a (001) heterostructure.
The octagon in Fig.~\ref{BZ} represents the projection of the bulk
Brillouin zone onto the (001) plane, and projections of sites
of the lattice are denoted by asterisks. The function
${\cal F}_n^{\prime}(k_z^{\prime}, {\bf k}_{||}^{\prime})$ on which the operator
${\cal M}_{nn^{\prime }}\left( {\bf k},{\bf k}^{\prime }\right)$ acts in Eq. (\ref{kp})
is defined only for ${\bf k}_{||}^{\prime}$ belonging to the projection of the
bulk Brillouin zone onto the (001) plane. But since all sites
${\bf K}_{||j}$ for $j\ne 0$ lie outside this projection, there exists a region
of ${\bf k}_{||}$ for which $\left| {\bf k}_{||}^{\prime}-{\bf k}_{||}\right|
<{\bf K}_{||j} $, $j\ne 0$. This region is defined by the inequality
$\left| k_x\right| +\left| k_y\right|<\pi /a$ and is indicated in the figure
by the square with diagonal $2\pi /a$. The area of this region is one-fourth that
of the area of the first 2D Brillouin zone (the square with diagonal $4\pi /a$).
In the region $\left| k_x\right| +\left| k_y\right|<\pi /a$ the second sum in expression
(\ref{sums}) does not contribute to the equations for the envelope functions. Its larger
dimensions ensure satisfaction of the conditions of applicability of the derived
equations for the envelope functions describing states near the ${\rm \Gamma }$ point.


\end{document}